\documentclass[
  aps,
  prl,
  twocolumn,
  superscriptaddress,
  floatfix,
  footinbib
]{revtex4-2}

\usepackage[utf8]{inputenc}
\usepackage[T1]{fontenc}
\usepackage{textcomp}
\usepackage{newunicodechar}
\newunicodechar{∘}{\textdegree}
\usepackage{lmodern}
\usepackage{amsmath,amssymb,amsthm}
\usepackage{physics}
\usepackage{graphicx}
\usepackage{hyperref}
\usepackage{xcolor}
\usepackage{braket}
\usepackage{bm}
\usepackage{bbm}
\usepackage{csquotes}

\hypersetup{
    colorlinks=true,
    linkcolor=blue,
    citecolor=blue,
    urlcolor=blue
}

\definecolor{revision}{RGB}{80, 160, 80}

\begin{document}

\title{Reversing the Chern Number with Quasiperiodic Disorder}

\author{B. D. K. Burgess}
\email{bb611@cam.ac.uk}
\affiliation{TCM Group, Cavendish Laboratory, University of Cambridge, J.J. Thomson Avenue, Cambridge, CB3 0US, United Kingdom\looseness=-1}

\author{N. R. Cooper}
\affiliation{TCM Group, Cavendish Laboratory, University of Cambridge, J.J. Thomson Avenue, Cambridge, CB3 0US, United Kingdom\looseness=-1}

\begin{abstract}
    We find an unconventional topological phase in the Haldane model driven by quasiperiodic disorder. Whereas increasing uncorrelated random disorder localizes states and destroys topology, increasing the amplitude of the specific quasiperiodic potential studied here confines extended states onto a network of `weak potential lines' (WPLs). By constructing an effective model for the system in this regime, we show that this confinement, along with virtual hopping processes via neighbouring sites, drives a topological transition into a phase with a reversed Chern number relative to the clean system. Our results demonstrate that quasiperiodic potentials can generate novel topological phases absent in both clean and randomly disordered systems.
\end{abstract}

\maketitle

Localisation of quantum particles has been a subject of interest in condensed matter physics since the original work of Anderson \cite{Anderson_Absence_1958}. In the absence of interactions, random disorder is found to localise all states for quantum systems in 1D and 2D \cite{Abrahams_Scaling_1979}, and extensive work on characterising localisation-delocalisation transitions has been performed \cite{Evers_Anderson_2008}. Localisation phenomena are also intimately connected to transitions in the quantum Hall effect \cite{Huckestein_Scaling_1995} and related topological insulators with broken time-reversal symmetry, with the precise universality classes and critical exponents the subject of ongoing research \cite{Slevin_Critical_2009, Amado_Numerical_2011, Nuding_Localization_2015, Gruzberg_Geometrically_2017, Klumper_Random_2019, Puschmann_Integer_2019, Zhu_Localizationlength_2019, Dresselhaus_Numerical_2021}. At strong disorder strengths, one generically finds a transition from topological to trivial phase via the `levitation and annihilation' mechanism \cite{Onoda_Localization_2007a, Prodan_Entanglement_2010, Prodan_Disordered_2011}. However, examples also exist where disorder can induce transitions into a topological phase, known as the topological Anderson insulator \cite{Li_Topological_2009, Groth_Theory_2009, Song_Dependence_2012, Garcia_RealSpace_2015, Orth_Topological_2016, Goncalves_Haldane_2018, Sriluckshmy_Interplay_2018, Vu_Weak_2022, Mildner_Topological_2023, Kovalenka_Universal_2026}. 

Quasiperiodic potentials provide an alternative setting for investigating localisation in quantum systems. Compared to random disorder, non-topological quasiperiodic systems can display various richer localisation properties, including mixed spectra where extended and localised states coexist \cite{Biddle_Predicted_2010, Biddle_Localization_2011, Ganeshan_Nearest_2015, Szabo_Mixed_2020, An_Interactions_2021}, intermediate `critically localised' states \cite{Zhu_Localization_2024, Duncan_Critical_2024}, and evidence of localisation transitions in different universality classes compared to random disorder \cite{Devakul_Anderson_2017, Szabo_Nonpowerlaw_2018}. There is also interest in quasicrystals for the potential of realising many-body localisation (MBL) in 2D \cite{Strkalj_Coexistence_2022, Thomson_Unravelling_2024, Sbroscia_Observing_2020, Hur_Stability_2025}, due to the absence of rare thermal regions which are theorised to destabilise MBL in 2D for random disorder \cite{DeRoeck_Stability_2017, Luitz_How_2017, Thiery_ManyBody_2018, Morningstar_Avalanches_2022}. Quasiperiodic potentials also naturally arise in both 2D layered materials \cite{Ahn_Dirac_2018, Yao_Quasicrystalline_2018, Yu_Dodecagonal_2019, Moon_Quasicrystalline_2019, Crosse_Quasicrystalline_2021, Mao_Quasiperiodicity_2021, Uri_Superconductivity_2023} and cold atom systems \cite{Roati_Anderson_2008, Schreiber_Observation_2015, Luschen_Observation_2017, Bordia_Coupling_2016, Bordia_Probing_2017, Viebahn_MatterWave_2019, Sbroscia_Observing_2020, Burgess_Quasicrystalline_2026, Neely_Programmable_2026}, making them of experimental relevance alongside inherent theoretical interest.

Given these examples of increased complexity compared to random disorder in non-topological contexts, it is natural to expect that quasiperiodic disorder can also give rise to novel effects in topological systems. Indeed, recent studies of such systems have described flat topological bands \cite{Fu_Flat_2021, Madeira_Quasidisorderinduced_2022}, unconventional phase transition mechanisms \cite{Goncalves_Topological_2022, Cheng_Topological_2023, Liquito_Fate_2024, Karcher_Extended_2024a}, and disorder-induced topological pumping \cite{Wu_Quantized_2022}. However, this area remains under-explored; in particular, the targeted investigation of specific quasiperiodic potentials known to host exotic localisation phenomena even in non-topological systems, offers a direct strategy for discovering topological phases with no counterpart in clean or randomly disordered systems. In this work, we study the Haldane model \cite{Haldane_Model_1988} subjected to one such potential with unconventional localisation behaviour: some states remain delocalised in the large potential limit \cite{Szabo_Mixed_2020, Strkalj_Coexistence_2022}. We will find that this potential can drive the system into a phase with opposite Chern number to the clean system, whose origin can be microscopically understood from these surviving extended states.

\begin{figure}[t]
    \centering
    \includegraphics[width=0.95\columnwidth]{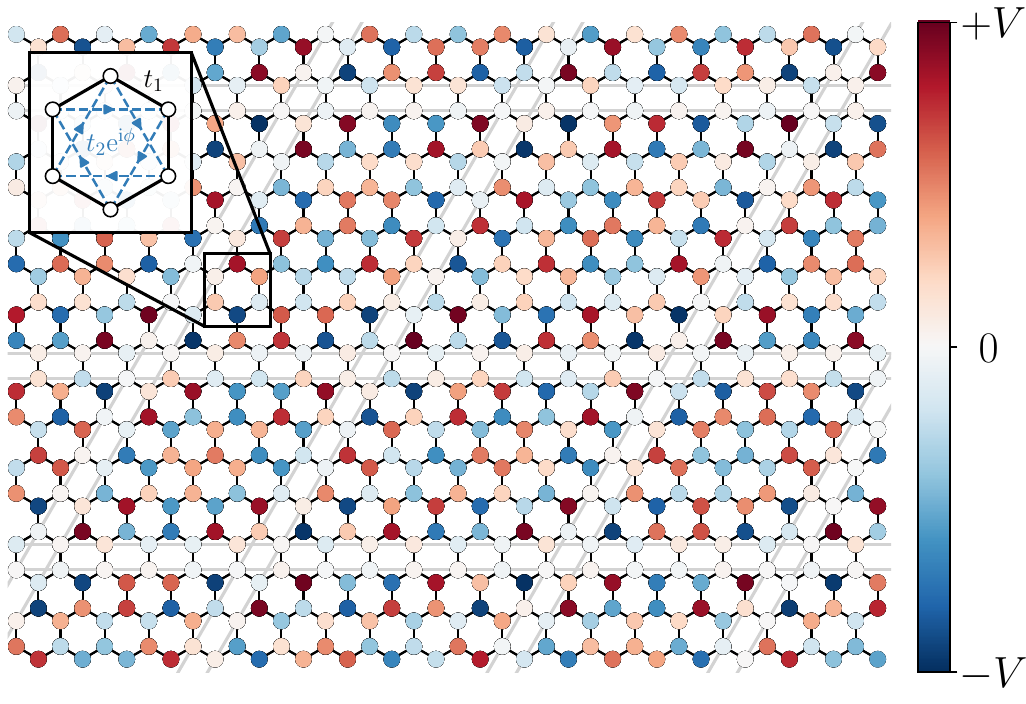}
    \caption{Section of honeycomb lattice showing the quasiperiodic on-site potential. WPLs are highlighted with grey lines. Inset: hoppings of the Haldane model. Arrows indicate the direction in which the NNN hopping is $t_2 \mathrm{e}^{\mathrm{i}\phi}$.}
    \label{Fig:Potential}
\end{figure}

\textit{Model---}%
We consider adding on-site disorder to the Haldane model \cite{Haldane_Model_1988} for spinless fermions on a honeycomb lattice. The Hamiltonian is given by
\begin{equation}\label{Eq:Hamiltonian}
\begin{aligned}
    H = & t_1 \! \left( \sum_{\langle ij \rangle} c_i^\dagger c_j + \mathrm{H.c.} \right)
    + t_2 \! \left( \sum_{\langle \langle ij \rangle \rangle} \mathrm{e}^{\mathrm{i}\phi_{ij}} c_i^\dagger c_j + \mathrm{H.c.} \right)\\
    + & M \! \left( \sum_{i \in  A} c_i^\dagger c_i 
    -\sum_{i \in  B} c_i^\dagger c_i \right) 
    + \sum_i V_i c_i^\dagger c_i \, .
\end{aligned}
\end{equation}
Here, $c_i^\dagger$ and $c_i$ are creation and annihilation operators at site $i$, and $\langle ij \rangle$ and $\langle \langle ij \rangle \rangle$ denote all pairs of nearest neighbours (NNs) and next-nearest neighbours (NNNs) respectively. $t_1$ is the NN hopping strength, and $t_2$ is the NNN hopping strength with complex phase $\phi_{ij}=+\phi$ in the directions indicated in Fig. \ref{Fig:Potential}, breaking time-reversal symmetry (TRS). We set $\phi=\pi/2$ throughout this work. The staggered mass $M$ breaks inversion symmetry between the two sublattices $A$ and $B$. $V_i$ is the disorder amplitude at site $i$. In the absence of disorder, the system is a Chern insulator with Chern number $C=\pm 1$ when $|3\sqrt{3}t_2\sin(\phi)|>|M|$.

In this work, we contrast the behaviour of the system when subjected to either random or quasiperiodic disorder. For random disorder, $V_i$ are drawn from a uniform distribution $V_i\in[-V, \, V]$. For quasiperiodic disorder, we utilise the potential
\begin{equation}\label{Eq:Potential}
    V_i = V\cos(2\pi\beta n + \theta_1)\cos(2 \pi \beta m + \theta_2),
\end{equation}
where $(n,m)$ are the coordinates of site $i$ in the basis of lattice vectors $\mathbf{a}_{1,2}$ (that is, $\mathbf{r}_i=n \mathbf{a}_1 + m \mathbf{a}_2$). $\beta\notin \mathbb{Q}$ is an irrational number which ensures the potential is incommensurate with the lattice, and hence quasiperiodic. Throughout the remainder of this work, we take $\beta=1/\sqrt{2}$ \footnote{We have found that potentials with other values of $\beta$ can similarly induce a reversal of Chern number, although generally following distinct mechanisms according to the microscopic details of the potential in each case. Likewise, a potential with an extra factor of $\cos(2\pi\beta(n+m))$, retaining the full 6-fold rotational symmetry of the honeycomb lattice, also drives a transition to $C=-1$. For simplicity of presentation, we focus on the potential of Eq. (\ref{Eq:Potential}) throughout this work.}. Varying the phases $\theta_{1,2}$ amounts to translating the origin of the potential, which enables us to sample different finite patches of the infinite system (i.e., different realisations of quasiperiodic disorder).

This potential is distinct from those employed in previous works investigating Chern insulators with quasiperiodic disorder \cite{Fu_Flat_2021, Goncalves_Topological_2022, Cheng_Topological_2023}. It possesses the special feature that, due to the quasiperiodicity, we are guaranteed to find lines of constant $m$ where $\cos(2\pi\beta m) \ll 1$. Therefore, particles may be delocalised in the $n$ direction even when $V\gg t_1, \, t_2$ (and vice versa for $n \leftrightarrow m$). These are termed weak potential lines (WPLs), and have been investigated in detail in \cite{Szabo_Mixed_2020} for single particles, and \cite{Strkalj_Coexistence_2022} in the case of many-body dynamics. Here, we investigate their interplay with topological properties in a TRS-broken system. 

While previous works examined square lattices \cite{Szabo_Mixed_2020, Strkalj_Coexistence_2022}, here we consider a honeycomb lattice. This introduces the additional feature that WPLs occur for neighbouring lines of $A$ and $B$ sites, resulting in the strips of low potential visible in Fig. \ref{Fig:Potential}. This is because the phase difference of the potential between these sites is $4\pi\beta/3\approx0.94 \pi$ for our choice of $\beta=1/\sqrt{2}$, so $\cos(2\pi\beta m) \ll 1$ along both lines. We also note that the localisation properties of the WPLs are primarily determined by the ratio $V/t_2$, since the WPLs consist of NNN bonds between sites on the same sublattice. For completeness, an explicit numerical verification of the eigenspectrum and localisation properties in this geometry is provided in the End Matter. Below, we turn directly to investigating the topological phase diagram of the system.

\textit{Topological properties---}%
To determine the topological character of the system in the presence of disorder, which breaks translation invariance, we calculate the real-space Chern marker \cite{Bianco_Mapping_2011}. This is defined locally at site $\mathbf{r}$ as
\begin{equation}
    c(\mathbf{r})=\frac{4\pi}{A_{\mathrm{c}}}\mathrm{Im} \big[ \bra{\mathbf{r}} P x P y P \ket{\mathbf{r}} \big] \, ,
\end{equation}
where $P=\sum_{n=1}^{N_{\mathrm{occ}}}\ket{\psi_n}\bra{\psi_n}$ is the projector onto occupied states and $A_{\mathrm{c}}$ is the unit-cell area. For periodic systems, the trace of $c(\mathbf{r})$ over a unit cell exactly equals the Chern number. For our calculations, we instead consider open boundary conditions on a finite honeycomb lattice of $L \times L$ unit cells ($2L^2$ total sites). We therefore characterise the topology with $\overline{c}$, the average Chern marker over the central $6 \times 6$ unit cells, deep within the bulk of the system. This approaches the integer-quantised Chern number value for sufficiently large $L$. We work at half-filling, $N_{\mathrm{occ}}=L^2$.

\begin{figure*}[t]
    \centering
    \includegraphics[width=\textwidth]{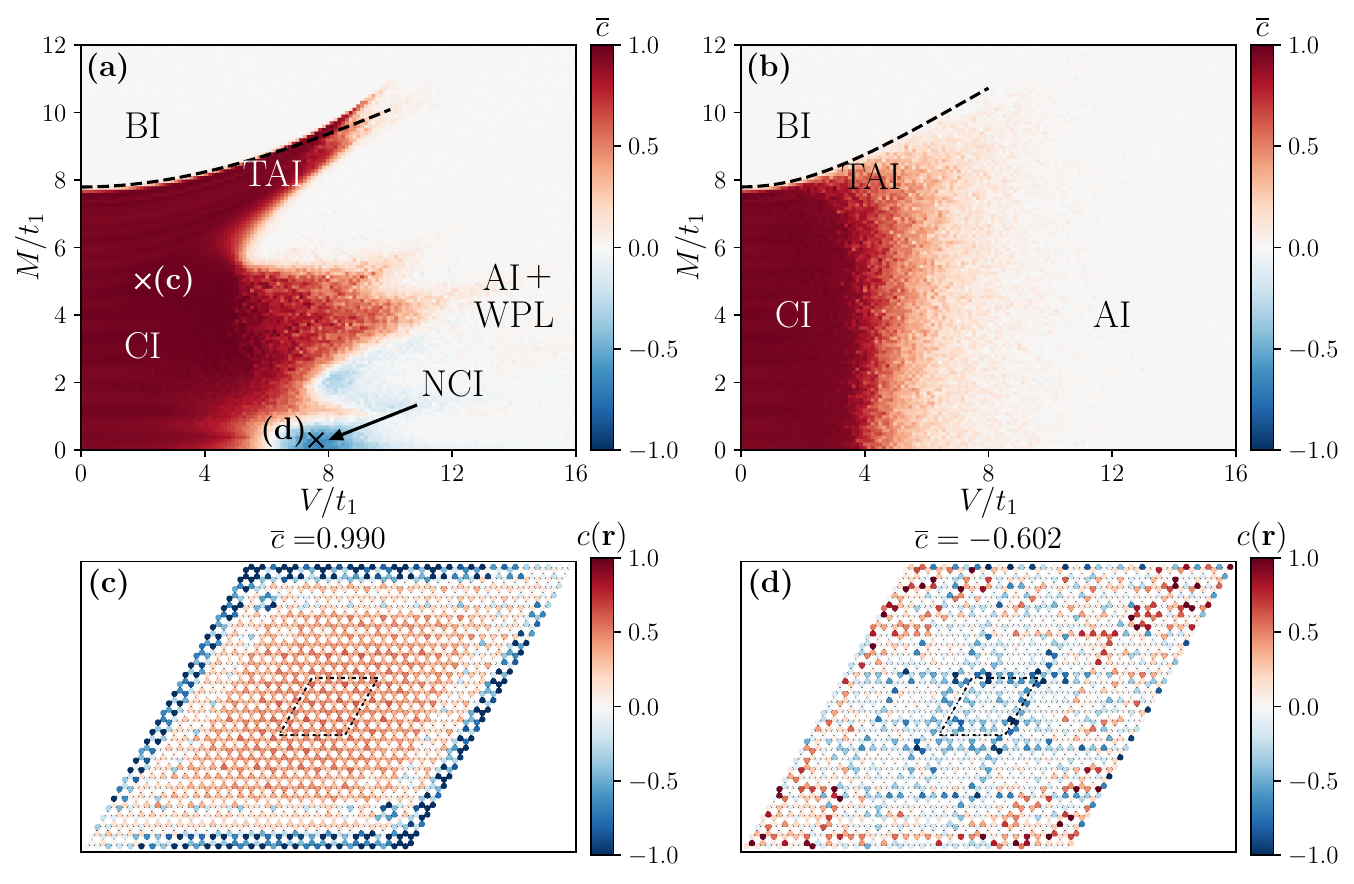}
    \caption{Phase diagram showing the bulk-averaged Chern marker of the Haldane model ($t_2/t_1=1.5$, $L=30$) as a function of disorder amplitude $V$ and on-site mass $M$, for (a) quasiperiodic and (b) random disorder. (Similar results for random disorder are presented in Refs.~\cite{Mildner_Topological_2023, Kovalenka_Universal_2026}.) Each point was calculated by averaging over 10 disorder configurations. Approximate regions of different phases (defined in the text) are labelled. Non-integer values of $\overline{c}$ are due to the finite $L$. Dashed lines show the phase boundaries calculated in the Supplemental Material \cite{SM}. Crosses in (a) indicate the parameters for which $c(\mathbf{r})$ is plotted in (c), (d). The dashed boxes in (c), (d) indicate the region over which $c(\mathbf{r})$ is averaged to obtain $\overline{c}$. The colour scales are capped at $\pm1$; values near the boundary can be significantly greater, since the sum of $c(\mathbf{r})$ over all sites must vanish in any finite system \cite{Bianco_Mapping_2011}.}
    \label{Fig:Phase_Diagram}
\end{figure*}

We plot phase diagrams of $\overline{c}$ as a function of disorder amplitude $V$ and mass $M$ in Fig. \ref{Fig:Phase_Diagram}, contrasting quasiperiodic and random disorder. Although the quasiperiodic diagram is noticeably more complex, we first highlight various features that are common to both cases. At low $V$ and $M$, the system resides in the clean Chern insulator (CI) phase, transitioning into a trivial band insulator (BI) above $M=3\sqrt{3}t_2$, and into a localised Anderson insulator (AI) at large $V$. In the quasiperiodic case the AI coexists with extended WPL states as detailed above (see also End Matter, Fig. \ref{Fig:Spectrum}). Both disorder types also exhibit a topological Anderson insulator (TAI) phase \cite{Li_Topological_2009}, where disorder drives the system from a trivial to a topological phase. This manifests in the upward curvature of the CI-BI boundary, and can be explained in terms of renormalisation of $M$ by the disorder \cite{Groth_Theory_2009}. Phase boundaries (dashed lines) derived within the Born approximation (see Supplemental Material \cite{SM}) agree well with the Chern marker at weak disorder.

In the system with quasiperiodic disorder, we find an additional phase with a negative Chern marker at intermediate disorder strengths, which we term the negative Chern insulator (NCI). This phase is caused by the WPLs. Extended states are necessary to support a non-trivial topology, and as $V$ increases, the only surviving extended states live along the WPLs. Accordingly, $c(\mathbf{r})$ becomes strongly concentrated along these paths [Fig. \ref{Fig:Phase_Diagram}(d)]. Below, we explain why this spatial confinement also reverses the sign of $c(\mathbf{r})$ relative to the clean system. In the Supplemental Material \cite{SM}, we further confirm that the NCI is a Chern insulator phase in the thermodynamic limit, with a robust bulk energy gap and quantized $\overline{c}=-1$ as $L \to \infty$.

\begin{figure}[t]
    \centering
    \includegraphics[width=\columnwidth]{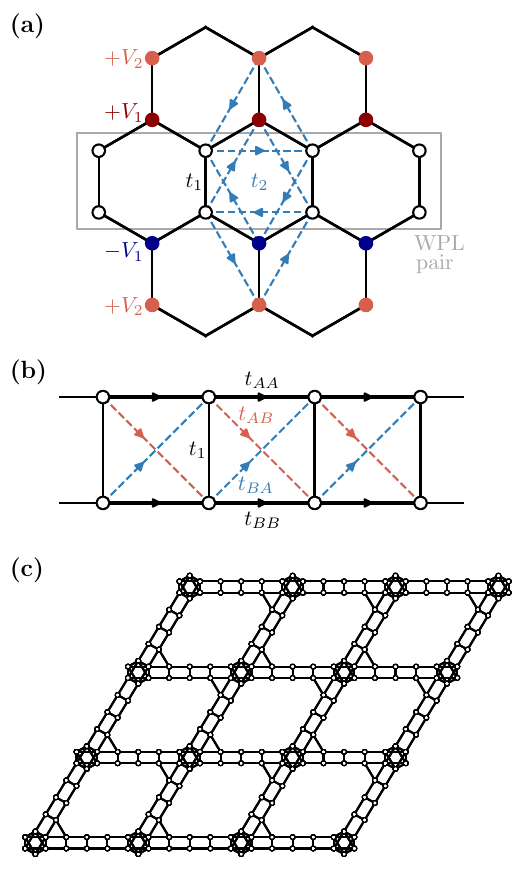}
    \caption{Construction of the effective ladder model. (a) Microscopic virtual hopping processes via neighbouring sites with on-site potentials $\pm V_1$ and $+V_2$. As in Fig. \ref{Fig:Potential}, arrows indicate the direction in which the NNN hopping amplitude is $t_2 \mathrm{e}^{+\mathrm{i}\phi}$. (b) Resulting ladder model with two sites per unit cell, labelled $A$ (upper) and $B$ (lower). Effective hopping elements, defined in Eq. (\ref{Eq:Ladder_Hopping_Elements}), are indicated. (c) Coupling of ladders into a 2D network. We retain NNN $t_2$ bonds connecting between adjacent ladders.}
    \label{Fig:Ladder_Construction}
\end{figure}

\textit{Effective model---}%
To gain insight into the physical origin of the NCI phase, we construct a simplified model for the system in this regime. The pairs of adjacent WPLs form a two-leg ladder. The effective hopping elements along the ladders contain contributions from both the bare hopping elements along each WPL, and also higher-order hopping processes via neighbouring sites, as illustrated in Fig. \ref{Fig:Ladder_Construction}(a). At second order (valid for sufficiently large $V$), we have
\begin{equation}\label{Eq:Effective_Hopping}
    t_{ij, \mathrm{eff}} = t_{ij}+\sum_l \frac{t_{il}t_{lj}}{-V_l} \, ,
\end{equation}
where $t_{ij}$ are the bare hopping matrix elements from site $i$ to $j$, the sum runs over all intermediate sites $l$ with on-site potential $V_l$, and we take $V_i=0$ for sites along the ladder.

In the full system, the amplitudes $V_l$ vary quasiperiodically from site to site. Here, we further simplify our model by considering the phase shift of the potential between the sites above and below the WPLs, which depends only on their vertical separation. For rows immediately above and below, the phase shift is $4\pi\beta/3\approx\pi$, so we set the potentials of these sites to $\pm V_1$, where $V_1\sim V$ is a representative average potential magnitude. For the rows one step further, the phase shift is $8\pi\beta/3\approx 2\pi$; we set both potentials to $V_2$. This simplification allows us to set the effective hopping elements equal along the whole ladder, as depicted in Fig.~\ref{Fig:Ladder_Construction}(b). Explicitly, with our choice $\phi=\pi/2$, the resulting effective hopping elements are:
\begin{equation}\label{Eq:Ladder_Hopping_Elements}
    \begin{aligned}
        t_{AA}&=\mathrm{i}t_2 - \frac{t_1^2}{V_1} - \frac{t_2^2}{V_1} + \frac{t_2^2}{V_2}\\
        t_{BB}&=-\mathrm{i}t_2 + \frac{t_1^2}{V_1} + \frac{t_2^2}{V_1} + \frac{t_2^2}{V_2}\\
        t_{AB}&=t_{BA}=-\frac{2\mathrm{i}t_1 t_2}{V_1} \, .
    \end{aligned}
\end{equation}
There is no renormalisation of the intra-cell $t_1$ hopping element due to cancellation of the virtual processes. We will find that our results are independent of the signs chosen for $V_1$ and $V_2$; only the relative signs between the two pairs of sites matter. We calculate explicit ($V$-dependent) values of $(V_1,V_2)$ by averaging over the relevant sites in a large finite system in the Supplemental Material \cite{SM}.

The final stage of our simplified model is to couple these ladders together into a 2D network [Fig. \ref{Fig:Ladder_Construction}(c)], mimicking the network of WPLs spanning the original system. We consider a network with ladders spaced by 5 lattice vectors. This approximates the quasiperiodic system, where $2\pi\beta \! \times \! 5\approx\pi \: (\mathrm{mod} \: 2\pi)$, so if row $m$ is a WPL (that is, $\cos (2\pi\beta m)\ll1$), then (usually) so is row $m+5$.

\begin{figure}[t]
    \centering
    \includegraphics[width=0.9\columnwidth]{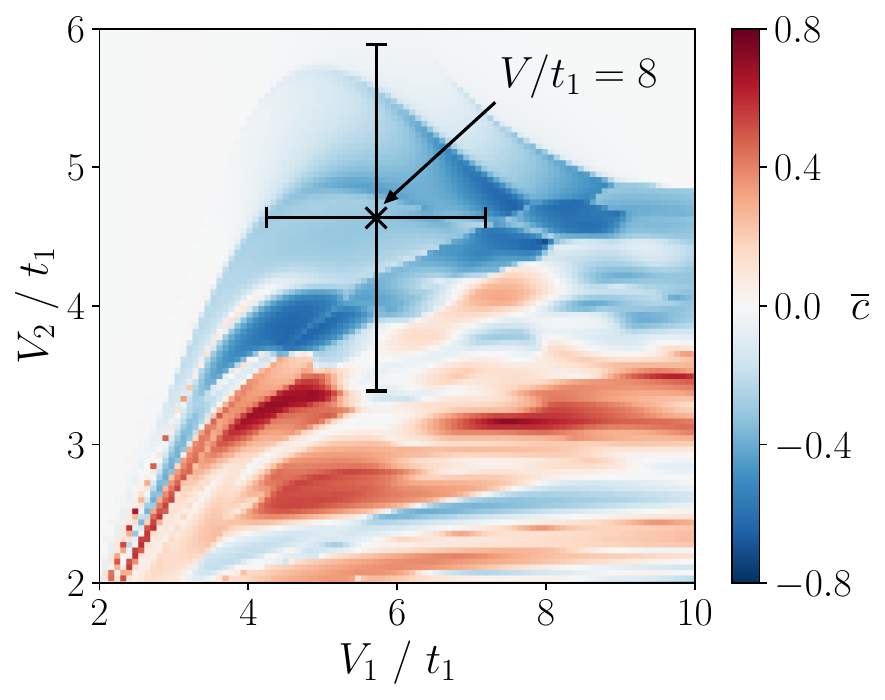}
    \caption{Phase diagram of the ladder network system ($t_2/t_1=1.5$, $L=37$). The Chern marker $\overline{c}$ is averaged over the central unit cell of the network (which has spatial periodicity $5a$). Because the unit cell size is significant relative to the system size, strong finite-size effects result in non-integer $\overline{c}$; nevertheless, the sign of $\overline{c}$ remains a reliable indicator of topology. The black cross and error bars denote the mean and standard deviation of $(V_1,V_2)$ across the system for $V/t_1=8$ \cite{SM}.}
    \label{Fig:Ladder_Phase_Diagram}
\end{figure}

The phase diagram of the ladder network [Fig.~\ref{Fig:Ladder_Phase_Diagram}] reveals a broad region with a negative Chern marker. Effective parameters $(V_1,V_2)$ calculated from the full quasiperiodic potential within the NCI regime $6 \lesssim V/t_1 \lesssim 10$ (see Supplemental Material \cite{SM}) fall within this domain, confirming that our simplified model captures the essential mechanism behind the reversal of Chern number. As a representative example, the values $(V_1,V_2)$ corresponding to $V/t_1=8$, in the middle of the NCI phase, are indicated by the black cross in Fig.~\ref{Fig:Ladder_Phase_Diagram}. Investigating different system sizes, we find that (unlike the NCI) our ladder model is metallic for these parameters, with vanishing energy gap in the limit $L\to\infty$ (see Supplemental Material \cite{SM}). Accordingly, $\overline{c}$ is not required to be integer-quantised; nevertheless, the sign of $\overline{c}$ enables us to make qualitative statements about the NCI phase.

In particular, the ladder model shows the role of the virtual hopping processes. At intermediate $(V_1, V_2)$ these are significant and give a negative $\overline{c}$. When $V_2$ is large they are suppressed, and $\overline{c}$ goes to zero. The same suppression in the full model explains why the NCI breaks down for $V/t_1 \gtrsim 10$. Note that the WPL states themselves remain extended up to much larger $V$ (see Fig.~\ref{Fig:Spectrum}), so this transition is not caused simply by their localization. The ladder model thus provides a complete microscopic picture, demonstrating that state confinement along WPLs together with virtual hoppings governs both the emergence of the NCI phase and its eventual termination at strong disorder.

\textit{Outlook---}%
We have shown that a quasiperiodic potential can reverse the Chern number of the Haldane model, leading to a NCI phase, through states confined to WPLs. Our results both illustrate that quasiperiodic potentials give rise to more complex behaviour than uncorrelated random disorder, and also show that the topological phase diagram depends on the detailed structure of the potential, not just its strength. This motivates future exploration of other quasiperiodic potentials with unusual localization properties as a way to find new topological phases.

Layered 2D materials provide a promising experimental platform where our model may be realised. The Haldane model has been experimentally realised in transition metal dichalcogenide (TMD) heterobilayers \cite{Zhao_Realization_2024}, and similarly emerges in the theoretical description of twisted TMD homobilayers \cite{Wu_Topological_2019}. At the same time, quasiperiodic phenomena naturally arise in these systems due to incommensurate twist angles or lattice spacings \cite{Ahn_Dirac_2018, Yao_Quasicrystalline_2018, Yu_Dodecagonal_2019, Moon_Quasicrystalline_2019, Crosse_Quasicrystalline_2021, Mao_Quasiperiodicity_2021, Uri_Superconductivity_2023}. A specific quasiperiodic potential could be induced in a Haldane layer, for example, through proximity to a twisted hexagonal boron nitride substrate \cite{Zhao_Universal_2021, Wang_Moire_2025}, allowing the tuning of both the potential wavelength and amplitude. We leave the detailed consideration of experimental requirements to future work.

\begin{acknowledgments}
\textit{Acknowledgments---}This work was supported by the EPSRC [Grants No. EP/V062654/1 and No. EP/Y01510X/1] and a Simons Investigator Award [Grant No. 511029]. 

\textit{Data and code availability---}The data and code supporting the findings of this article are openly available \cite{Data}.
\end{acknowledgments}

\bibliography{refs}

\begin{thebibliography}{74}%
\makeatletter
\providecommand \@ifxundefined [1]{%
 \@ifx{#1\undefined}
}%
\providecommand \@ifnum [1]{%
 \ifnum #1\expandafter \@firstoftwo
 \else \expandafter \@secondoftwo
 \fi
}%
\providecommand \@ifx [1]{%
 \ifx #1\expandafter \@firstoftwo
 \else \expandafter \@secondoftwo
 \fi
}%
\providecommand \natexlab [1]{#1}%
\providecommand \enquote  [1]{``#1''}%
\providecommand \bibnamefont  [1]{#1}%
\providecommand \bibfnamefont [1]{#1}%
\providecommand \citenamefont [1]{#1}%
\providecommand \href@noop [0]{\@secondoftwo}%
\providecommand \href [0]{\begingroup \@sanitize@url \@href}%
\providecommand \@href[1]{\@@startlink{#1}\@@href}%
\providecommand \@@href[1]{\endgroup#1\@@endlink}%
\providecommand \@sanitize@url [0]{\catcode `\\12\catcode `\$12\catcode
  `\&12\catcode `\#12\catcode `\^12\catcode `\_12\catcode `\%12\relax}%
\providecommand \@@startlink[1]{}%
\providecommand \@@endlink[0]{}%
\providecommand \url  [0]{\begingroup\@sanitize@url \@url }%
\providecommand \@url [1]{\endgroup\@href {#1}{\urlprefix }}%
\providecommand \urlprefix  [0]{URL }%
\providecommand \Eprint [0]{\href }%
\providecommand \doibase [0]{https://doi.org/}%
\providecommand \selectlanguage [0]{\@gobble}%
\providecommand \bibinfo  [0]{\@secondoftwo}%
\providecommand \bibfield  [0]{\@secondoftwo}%
\providecommand \translation [1]{[#1]}%
\providecommand \BibitemOpen [0]{}%
\providecommand \bibitemStop [0]{}%
\providecommand \bibitemNoStop [0]{.\EOS\space}%
\providecommand \EOS [0]{\spacefactor3000\relax}%
\providecommand \BibitemShut  [1]{\csname bibitem#1\endcsname}%
\let\auto@bib@innerbib\@empty
\bibitem [{\citenamefont {Anderson}(1958)}]{Anderson_Absence_1958}%
  \BibitemOpen
  \bibfield  {author} {\bibinfo {author} {\bibfnamefont {P.}~\bibnamefont
  {Anderson}},\ }\bibfield  {title} {\bibinfo {title} {Absence of diffusion in
  certain random lattices},\ }\href@noop {} {\bibfield  {journal} {\bibinfo
  {journal} {Physical review}\ }\textbf {\bibinfo {volume} {109}},\ \bibinfo
  {pages} {1492} (\bibinfo {year} {1958})}\BibitemShut {NoStop}%
\bibitem [{\citenamefont {Abrahams}\ \emph {et~al.}(1979)\citenamefont
  {Abrahams}, \citenamefont {Anderson}, \citenamefont {Licciardello},\ and\
  \citenamefont {Ramakrishnan}}]{Abrahams_Scaling_1979}%
  \BibitemOpen
  \bibfield  {author} {\bibinfo {author} {\bibfnamefont {E.}~\bibnamefont
  {Abrahams}}, \bibinfo {author} {\bibfnamefont {P.~W.}\ \bibnamefont
  {Anderson}}, \bibinfo {author} {\bibfnamefont {D.~C.}\ \bibnamefont
  {Licciardello}},\ and\ \bibinfo {author} {\bibfnamefont {T.~V.}\ \bibnamefont
  {Ramakrishnan}},\ }\bibfield  {title} {\bibinfo {title} {Scaling {{Theory}}
  of {{Localization}}: {{Absence}} of {{Quantum Diffusion}} in {{Two
  Dimensions}}},\ }\href {https://doi.org/10.1103/PhysRevLett.42.673}
  {\bibfield  {journal} {\bibinfo  {journal} {Physical Review Letters}\
  }\textbf {\bibinfo {volume} {42}},\ \bibinfo {pages} {673} (\bibinfo {year}
  {1979})}\BibitemShut {NoStop}%
\bibitem [{\citenamefont {Evers}\ and\ \citenamefont
  {Mirlin}(2008)}]{Evers_Anderson_2008}%
  \BibitemOpen
  \bibfield  {author} {\bibinfo {author} {\bibfnamefont {F.}~\bibnamefont
  {Evers}}\ and\ \bibinfo {author} {\bibfnamefont {A.~D.}\ \bibnamefont
  {Mirlin}},\ }\bibfield  {title} {\bibinfo {title} {Anderson transitions},\
  }\href {https://doi.org/10.1103/RevModPhys.80.1355} {\bibfield  {journal}
  {\bibinfo  {journal} {Reviews of Modern Physics}\ }\textbf {\bibinfo {volume}
  {80}},\ \bibinfo {pages} {1355} (\bibinfo {year} {2008})}\BibitemShut
  {NoStop}%
\bibitem [{\citenamefont {Huckestein}(1995)}]{Huckestein_Scaling_1995}%
  \BibitemOpen
  \bibfield  {author} {\bibinfo {author} {\bibfnamefont {B.}~\bibnamefont
  {Huckestein}},\ }\bibfield  {title} {\bibinfo {title} {Scaling theory of the
  integer quantum {{Hall}} effect},\ }\href
  {https://doi.org/10.1103/RevModPhys.67.357} {\bibfield  {journal} {\bibinfo
  {journal} {Reviews of Modern Physics}\ }\textbf {\bibinfo {volume} {67}},\
  \bibinfo {pages} {357} (\bibinfo {year} {1995})}\BibitemShut {NoStop}%
\bibitem [{\citenamefont {Slevin}\ and\ \citenamefont
  {Ohtsuki}(2009)}]{Slevin_Critical_2009}%
  \BibitemOpen
  \bibfield  {author} {\bibinfo {author} {\bibfnamefont {K.}~\bibnamefont
  {Slevin}}\ and\ \bibinfo {author} {\bibfnamefont {T.}~\bibnamefont
  {Ohtsuki}},\ }\bibfield  {title} {\bibinfo {title} {Critical exponent for the
  quantum {{Hall}} transition},\ }\href
  {https://doi.org/10.1103/PhysRevB.80.041304} {\bibfield  {journal} {\bibinfo
  {journal} {Physical Review B}\ }\textbf {\bibinfo {volume} {80}},\ \bibinfo
  {pages} {041304} (\bibinfo {year} {2009})}\BibitemShut {NoStop}%
\bibitem [{\citenamefont {Amado}\ \emph {et~al.}(2011)\citenamefont {Amado},
  \citenamefont {Malyshev}, \citenamefont {Sedrakyan},\ and\ \citenamefont
  {{Dom{\'i}nguez-Adame}}}]{Amado_Numerical_2011}%
  \BibitemOpen
  \bibfield  {author} {\bibinfo {author} {\bibfnamefont {M.}~\bibnamefont
  {Amado}}, \bibinfo {author} {\bibfnamefont {A.~V.}\ \bibnamefont {Malyshev}},
  \bibinfo {author} {\bibfnamefont {A.}~\bibnamefont {Sedrakyan}},\ and\
  \bibinfo {author} {\bibfnamefont {F.}~\bibnamefont {{Dom{\'i}nguez-Adame}}},\
  }\bibfield  {title} {\bibinfo {title} {Numerical {{Study}} of the
  {{Localization Length Critical Index}} in a {{Network Model}} of
  {{Plateau-Plateau Transitions}} in the {{Quantum Hall Effect}}},\ }\href
  {https://doi.org/10.1103/PhysRevLett.107.066402} {\bibfield  {journal}
  {\bibinfo  {journal} {Physical Review Letters}\ }\textbf {\bibinfo {volume}
  {107}},\ \bibinfo {pages} {066402} (\bibinfo {year} {2011})}\BibitemShut
  {NoStop}%
\bibitem [{\citenamefont {Nuding}\ \emph {et~al.}(2015)\citenamefont {Nuding},
  \citenamefont {Kl{\"u}mper},\ and\ \citenamefont
  {Sedrakyan}}]{Nuding_Localization_2015}%
  \BibitemOpen
  \bibfield  {author} {\bibinfo {author} {\bibfnamefont {W.}~\bibnamefont
  {Nuding}}, \bibinfo {author} {\bibfnamefont {A.}~\bibnamefont
  {Kl{\"u}mper}},\ and\ \bibinfo {author} {\bibfnamefont {A.}~\bibnamefont
  {Sedrakyan}},\ }\bibfield  {title} {\bibinfo {title} {Localization length
  index and subleading corrections in a {{Chalker-Coddington}} model: {{A}}
  numerical study},\ }\href {https://doi.org/10.1103/PhysRevB.91.115107}
  {\bibfield  {journal} {\bibinfo  {journal} {Physical Review B}\ }\textbf
  {\bibinfo {volume} {91}},\ \bibinfo {pages} {115107} (\bibinfo {year}
  {2015})}\BibitemShut {NoStop}%
\bibitem [{\citenamefont {Gruzberg}\ \emph {et~al.}(2017)\citenamefont
  {Gruzberg}, \citenamefont {Kl{\"u}mper}, \citenamefont {Nuding},\ and\
  \citenamefont {Sedrakyan}}]{Gruzberg_Geometrically_2017}%
  \BibitemOpen
  \bibfield  {author} {\bibinfo {author} {\bibfnamefont {I.~A.}\ \bibnamefont
  {Gruzberg}}, \bibinfo {author} {\bibfnamefont {A.}~\bibnamefont
  {Kl{\"u}mper}}, \bibinfo {author} {\bibfnamefont {W.}~\bibnamefont
  {Nuding}},\ and\ \bibinfo {author} {\bibfnamefont {A.}~\bibnamefont
  {Sedrakyan}},\ }\bibfield  {title} {\bibinfo {title} {Geometrically
  disordered network models, quenched quantum gravity, and critical behavior at
  quantum {{Hall}} plateau transitions},\ }\href
  {https://doi.org/10.1103/PhysRevB.95.125414} {\bibfield  {journal} {\bibinfo
  {journal} {Physical Review B}\ }\textbf {\bibinfo {volume} {95}},\ \bibinfo
  {pages} {125414} (\bibinfo {year} {2017})}\BibitemShut {NoStop}%
\bibitem [{\citenamefont {Kl{\"u}mper}\ \emph {et~al.}(2019)\citenamefont
  {Kl{\"u}mper}, \citenamefont {Nuding},\ and\ \citenamefont
  {Sedrakyan}}]{Klumper_Random_2019}%
  \BibitemOpen
  \bibfield  {author} {\bibinfo {author} {\bibfnamefont {A.}~\bibnamefont
  {Kl{\"u}mper}}, \bibinfo {author} {\bibfnamefont {W.}~\bibnamefont
  {Nuding}},\ and\ \bibinfo {author} {\bibfnamefont {A.}~\bibnamefont
  {Sedrakyan}},\ }\bibfield  {title} {\bibinfo {title} {Random network models
  with variable disorder of geometry},\ }\href
  {https://doi.org/10.1103/PhysRevB.100.140201} {\bibfield  {journal} {\bibinfo
   {journal} {Physical Review B}\ }\textbf {\bibinfo {volume} {100}},\ \bibinfo
  {pages} {140201} (\bibinfo {year} {2019})}\BibitemShut {NoStop}%
\bibitem [{\citenamefont {Puschmann}\ \emph {et~al.}(2019)\citenamefont
  {Puschmann}, \citenamefont {Cain}, \citenamefont {Schreiber},\ and\
  \citenamefont {Vojta}}]{Puschmann_Integer_2019}%
  \BibitemOpen
  \bibfield  {author} {\bibinfo {author} {\bibfnamefont {M.}~\bibnamefont
  {Puschmann}}, \bibinfo {author} {\bibfnamefont {P.}~\bibnamefont {Cain}},
  \bibinfo {author} {\bibfnamefont {M.}~\bibnamefont {Schreiber}},\ and\
  \bibinfo {author} {\bibfnamefont {T.}~\bibnamefont {Vojta}},\ }\bibfield
  {title} {\bibinfo {title} {Integer quantum {{Hall}} transition on a
  tight-binding lattice},\ }\href {https://doi.org/10.1103/PhysRevB.99.121301}
  {\bibfield  {journal} {\bibinfo  {journal} {Physical Review B}\ }\textbf
  {\bibinfo {volume} {99}},\ \bibinfo {pages} {121301} (\bibinfo {year}
  {2019})}\BibitemShut {NoStop}%
\bibitem [{\citenamefont {Zhu}\ \emph {et~al.}(2019)\citenamefont {Zhu},
  \citenamefont {Wu}, \citenamefont {Bhatt},\ and\ \citenamefont
  {Wan}}]{Zhu_Localizationlength_2019}%
  \BibitemOpen
  \bibfield  {author} {\bibinfo {author} {\bibfnamefont {Q.}~\bibnamefont
  {Zhu}}, \bibinfo {author} {\bibfnamefont {P.}~\bibnamefont {Wu}}, \bibinfo
  {author} {\bibfnamefont {R.~N.}\ \bibnamefont {Bhatt}},\ and\ \bibinfo
  {author} {\bibfnamefont {X.}~\bibnamefont {Wan}},\ }\bibfield  {title}
  {\bibinfo {title} {Localization-length exponent in two models of quantum
  {{Hall}} plateau transitions},\ }\href
  {https://doi.org/10.1103/PhysRevB.99.024205} {\bibfield  {journal} {\bibinfo
  {journal} {Physical Review B}\ }\textbf {\bibinfo {volume} {99}},\ \bibinfo
  {pages} {024205} (\bibinfo {year} {2019})}\BibitemShut {NoStop}%
\bibitem [{\citenamefont {Dresselhaus}\ \emph {et~al.}(2021)\citenamefont
  {Dresselhaus}, \citenamefont {Sbierski},\ and\ \citenamefont
  {Gruzberg}}]{Dresselhaus_Numerical_2021}%
  \BibitemOpen
  \bibfield  {author} {\bibinfo {author} {\bibfnamefont {E.~J.}\ \bibnamefont
  {Dresselhaus}}, \bibinfo {author} {\bibfnamefont {B.}~\bibnamefont
  {Sbierski}},\ and\ \bibinfo {author} {\bibfnamefont {I.~A.}\ \bibnamefont
  {Gruzberg}},\ }\bibfield  {title} {\bibinfo {title} {Numerical evidence for
  marginal scaling at the integer quantum {{Hall}} transition},\ }\href
  {https://doi.org/10.1016/j.aop.2021.168676} {\bibfield  {journal} {\bibinfo
  {journal} {Annals of Physics}\ }\bibinfo {series} {Special {{Issue}} on
  {{Localisation}} 2020},\ \textbf {\bibinfo {volume} {435}},\ \bibinfo {pages}
  {168676} (\bibinfo {year} {2021})}\BibitemShut {NoStop}%
\bibitem [{\citenamefont {Onoda}\ \emph {et~al.}(2007)\citenamefont {Onoda},
  \citenamefont {Avishai},\ and\ \citenamefont
  {Nagaosa}}]{Onoda_Localization_2007a}%
  \BibitemOpen
  \bibfield  {author} {\bibinfo {author} {\bibfnamefont {M.}~\bibnamefont
  {Onoda}}, \bibinfo {author} {\bibfnamefont {Y.}~\bibnamefont {Avishai}},\
  and\ \bibinfo {author} {\bibfnamefont {N.}~\bibnamefont {Nagaosa}},\
  }\bibfield  {title} {\bibinfo {title} {Localization in a {{Quantum Spin Hall
  System}}},\ }\href {https://doi.org/10.1103/PhysRevLett.98.076802} {\bibfield
   {journal} {\bibinfo  {journal} {Physical Review Letters}\ }\textbf {\bibinfo
  {volume} {98}},\ \bibinfo {pages} {076802} (\bibinfo {year}
  {2007})}\BibitemShut {NoStop}%
\bibitem [{\citenamefont {Prodan}\ \emph {et~al.}(2010)\citenamefont {Prodan},
  \citenamefont {Hughes},\ and\ \citenamefont
  {Bernevig}}]{Prodan_Entanglement_2010}%
  \BibitemOpen
  \bibfield  {author} {\bibinfo {author} {\bibfnamefont {E.}~\bibnamefont
  {Prodan}}, \bibinfo {author} {\bibfnamefont {T.~L.}\ \bibnamefont {Hughes}},\
  and\ \bibinfo {author} {\bibfnamefont {B.~A.}\ \bibnamefont {Bernevig}},\
  }\bibfield  {title} {\bibinfo {title} {Entanglement {{Spectrum}} of a
  {{Disordered Topological Chern Insulator}}},\ }\href
  {https://doi.org/10.1103/PhysRevLett.105.115501} {\bibfield  {journal}
  {\bibinfo  {journal} {Physical Review Letters}\ }\textbf {\bibinfo {volume}
  {105}},\ \bibinfo {pages} {115501} (\bibinfo {year} {2010})}\BibitemShut
  {NoStop}%
\bibitem [{\citenamefont {Prodan}(2011)}]{Prodan_Disordered_2011}%
  \BibitemOpen
  \bibfield  {author} {\bibinfo {author} {\bibfnamefont {E.}~\bibnamefont
  {Prodan}},\ }\bibfield  {title} {\bibinfo {title} {Disordered topological
  insulators: A non-commutative geometry perspective},\ }\href
  {https://doi.org/10.1088/1751-8113/44/11/113001} {\bibfield  {journal}
  {\bibinfo  {journal} {Journal of Physics A: Mathematical and Theoretical}\
  }\textbf {\bibinfo {volume} {44}},\ \bibinfo {pages} {113001} (\bibinfo
  {year} {2011})}\BibitemShut {NoStop}%
\bibitem [{\citenamefont {Li}\ \emph {et~al.}(2009)\citenamefont {Li},
  \citenamefont {Chu}, \citenamefont {Jain},\ and\ \citenamefont
  {Shen}}]{Li_Topological_2009}%
  \BibitemOpen
  \bibfield  {author} {\bibinfo {author} {\bibfnamefont {J.}~\bibnamefont
  {Li}}, \bibinfo {author} {\bibfnamefont {R.-L.}\ \bibnamefont {Chu}},
  \bibinfo {author} {\bibfnamefont {J.~K.}\ \bibnamefont {Jain}},\ and\
  \bibinfo {author} {\bibfnamefont {S.-Q.}\ \bibnamefont {Shen}},\ }\bibfield
  {title} {\bibinfo {title} {Topological {{Anderson Insulator}}},\ }\href
  {https://doi.org/10.1103/PhysRevLett.102.136806} {\bibfield  {journal}
  {\bibinfo  {journal} {Physical Review Letters}\ }\textbf {\bibinfo {volume}
  {102}},\ \bibinfo {pages} {136806} (\bibinfo {year} {2009})}\BibitemShut
  {NoStop}%
\bibitem [{\citenamefont {Groth}\ \emph {et~al.}(2009)\citenamefont {Groth},
  \citenamefont {Wimmer}, \citenamefont {Akhmerov}, \citenamefont
  {Tworzyd{\l}o},\ and\ \citenamefont {Beenakker}}]{Groth_Theory_2009}%
  \BibitemOpen
  \bibfield  {author} {\bibinfo {author} {\bibfnamefont {C.~W.}\ \bibnamefont
  {Groth}}, \bibinfo {author} {\bibfnamefont {M.}~\bibnamefont {Wimmer}},
  \bibinfo {author} {\bibfnamefont {A.~R.}\ \bibnamefont {Akhmerov}}, \bibinfo
  {author} {\bibfnamefont {J.}~\bibnamefont {Tworzyd{\l}o}},\ and\ \bibinfo
  {author} {\bibfnamefont {C.~W.~J.}\ \bibnamefont {Beenakker}},\ }\bibfield
  {title} {\bibinfo {title} {Theory of the {{Topological Anderson
  Insulator}}},\ }\href {https://doi.org/10.1103/PhysRevLett.103.196805}
  {\bibfield  {journal} {\bibinfo  {journal} {Physical Review Letters}\
  }\textbf {\bibinfo {volume} {103}},\ \bibinfo {pages} {196805} (\bibinfo
  {year} {2009})}\BibitemShut {NoStop}%
\bibitem [{\citenamefont {Song}\ \emph {et~al.}(2012)\citenamefont {Song},
  \citenamefont {Liu}, \citenamefont {Jiang}, \citenamefont {Sun},\ and\
  \citenamefont {Xie}}]{Song_Dependence_2012}%
  \BibitemOpen
  \bibfield  {author} {\bibinfo {author} {\bibfnamefont {J.}~\bibnamefont
  {Song}}, \bibinfo {author} {\bibfnamefont {H.}~\bibnamefont {Liu}}, \bibinfo
  {author} {\bibfnamefont {H.}~\bibnamefont {Jiang}}, \bibinfo {author}
  {\bibfnamefont {Q.-f.}\ \bibnamefont {Sun}},\ and\ \bibinfo {author}
  {\bibfnamefont {X.~C.}\ \bibnamefont {Xie}},\ }\bibfield  {title} {\bibinfo
  {title} {Dependence of topological {{Anderson}} insulator on the type of
  disorder},\ }\href {https://doi.org/10.1103/PhysRevB.85.195125} {\bibfield
  {journal} {\bibinfo  {journal} {Physical Review B}\ }\textbf {\bibinfo
  {volume} {85}},\ \bibinfo {pages} {195125} (\bibinfo {year}
  {2012})}\BibitemShut {NoStop}%
\bibitem [{\citenamefont {Garc{\'i}a}\ \emph {et~al.}(2015)\citenamefont
  {Garc{\'i}a}, \citenamefont {Covaci},\ and\ \citenamefont
  {Rappoport}}]{Garcia_RealSpace_2015}%
  \BibitemOpen
  \bibfield  {author} {\bibinfo {author} {\bibfnamefont {J.~H.}\ \bibnamefont
  {Garc{\'i}a}}, \bibinfo {author} {\bibfnamefont {L.}~\bibnamefont {Covaci}},\
  and\ \bibinfo {author} {\bibfnamefont {T.~G.}\ \bibnamefont {Rappoport}},\
  }\bibfield  {title} {\bibinfo {title} {Real-{{Space Calculation}} of the
  {{Conductivity Tensor}} for {{Disordered Topological Matter}}},\ }\href
  {https://doi.org/10.1103/PhysRevLett.114.116602} {\bibfield  {journal}
  {\bibinfo  {journal} {Physical Review Letters}\ }\textbf {\bibinfo {volume}
  {114}},\ \bibinfo {pages} {116602} (\bibinfo {year} {2015})}\BibitemShut
  {NoStop}%
\bibitem [{\citenamefont {Orth}\ \emph {et~al.}(2016)\citenamefont {Orth},
  \citenamefont {Sekera}, \citenamefont {Bruder},\ and\ \citenamefont
  {Schmidt}}]{Orth_Topological_2016}%
  \BibitemOpen
  \bibfield  {author} {\bibinfo {author} {\bibfnamefont {C.~P.}\ \bibnamefont
  {Orth}}, \bibinfo {author} {\bibfnamefont {T.}~\bibnamefont {Sekera}},
  \bibinfo {author} {\bibfnamefont {C.}~\bibnamefont {Bruder}},\ and\ \bibinfo
  {author} {\bibfnamefont {T.~L.}\ \bibnamefont {Schmidt}},\ }\bibfield
  {title} {\bibinfo {title} {The topological {{Anderson}} insulator phase in
  the {{Kane-Mele}} model},\ }\href {https://doi.org/10.1038/srep24007}
  {\bibfield  {journal} {\bibinfo  {journal} {Scientific Reports}\ }\textbf
  {\bibinfo {volume} {6}},\ \bibinfo {pages} {24007} (\bibinfo {year}
  {2016})}\BibitemShut {NoStop}%
\bibitem [{\citenamefont {Gon{\c c}alves}\ \emph {et~al.}(2018)\citenamefont
  {Gon{\c c}alves}, \citenamefont {Ribeiro},\ and\ \citenamefont
  {Castro}}]{Goncalves_Haldane_2018}%
  \BibitemOpen
  \bibfield  {author} {\bibinfo {author} {\bibfnamefont {M.}~\bibnamefont
  {Gon{\c c}alves}}, \bibinfo {author} {\bibfnamefont {P.}~\bibnamefont
  {Ribeiro}},\ and\ \bibinfo {author} {\bibfnamefont {E.~V.}\ \bibnamefont
  {Castro}},\ }\href {https://doi.org/10.48550/arXiv.1807.11247} {\bibinfo
  {title} {The {{Haldane}} model under quenched disorder}} (\bibinfo {year}
  {2018}),\ \Eprint {https://arxiv.org/abs/1807.11247} {arXiv:1807.11247
  [cond-mat.str-el]} \BibitemShut {NoStop}%
\bibitem [{\citenamefont {Sriluckshmy}\ \emph {et~al.}(2018)\citenamefont
  {Sriluckshmy}, \citenamefont {Saha},\ and\ \citenamefont
  {Moessner}}]{Sriluckshmy_Interplay_2018}%
  \BibitemOpen
  \bibfield  {author} {\bibinfo {author} {\bibfnamefont {P.~V.}\ \bibnamefont
  {Sriluckshmy}}, \bibinfo {author} {\bibfnamefont {K.}~\bibnamefont {Saha}},\
  and\ \bibinfo {author} {\bibfnamefont {R.}~\bibnamefont {Moessner}},\
  }\bibfield  {title} {\bibinfo {title} {Interplay between topology and
  disorder in a two-dimensional semi-{{Dirac}} material},\ }\href
  {https://doi.org/10.1103/PhysRevB.97.024204} {\bibfield  {journal} {\bibinfo
  {journal} {Physical Review B}\ }\textbf {\bibinfo {volume} {97}},\ \bibinfo
  {pages} {024204} (\bibinfo {year} {2018})}\BibitemShut {NoStop}%
\bibitem [{\citenamefont {Vu}\ and\ \citenamefont
  {Das~Sarma}(2022)}]{Vu_Weak_2022}%
  \BibitemOpen
  \bibfield  {author} {\bibinfo {author} {\bibfnamefont {D.}~\bibnamefont
  {Vu}}\ and\ \bibinfo {author} {\bibfnamefont {S.}~\bibnamefont {Das~Sarma}},\
  }\bibfield  {title} {\bibinfo {title} {Weak quantization of noninteracting
  topological {{Anderson}} insulator},\ }\href
  {https://doi.org/10.1103/PhysRevB.106.134201} {\bibfield  {journal} {\bibinfo
   {journal} {Physical Review B}\ }\textbf {\bibinfo {volume} {106}},\ \bibinfo
  {pages} {134201} (\bibinfo {year} {2022})}\BibitemShut {NoStop}%
\bibitem [{\citenamefont {Mildner}\ \emph {et~al.}(2023)\citenamefont
  {Mildner}, \citenamefont {Caio}, \citenamefont {M{\"o}ller}, \citenamefont
  {Cooper},\ and\ \citenamefont {Bhaseen}}]{Mildner_Topological_2023}%
  \BibitemOpen
  \bibfield  {author} {\bibinfo {author} {\bibfnamefont {J.}~\bibnamefont
  {Mildner}}, \bibinfo {author} {\bibfnamefont {M.~D.}\ \bibnamefont {Caio}},
  \bibinfo {author} {\bibfnamefont {G.}~\bibnamefont {M{\"o}ller}}, \bibinfo
  {author} {\bibfnamefont {N.~R.}\ \bibnamefont {Cooper}},\ and\ \bibinfo
  {author} {\bibfnamefont {M.~J.}\ \bibnamefont {Bhaseen}},\ }\href
  {https://doi.org/10.48550/arXiv.2312.16689} {\bibinfo {title} {Topological
  {{Phase Transitions}} in the {{Disordered Haldane Model}}}} (\bibinfo {year}
  {2023}),\ \Eprint {https://arxiv.org/abs/2312.16689} {arXiv:2312.16689
  [cond-mat]} \BibitemShut {NoStop}%
\bibitem [{\citenamefont {Kovalenka}\ \emph {et~al.}(2026)\citenamefont
  {Kovalenka}, \citenamefont {Ranjbar}, \citenamefont {Azadi}, \citenamefont
  {Belosludov}, \citenamefont {K{\"u}hne},\ and\ \citenamefont
  {Bahramy}}]{Kovalenka_Universal_2026}%
  \BibitemOpen
  \bibfield  {author} {\bibinfo {author} {\bibfnamefont {K.}~\bibnamefont
  {Kovalenka}}, \bibinfo {author} {\bibfnamefont {A.}~\bibnamefont {Ranjbar}},
  \bibinfo {author} {\bibfnamefont {S.}~\bibnamefont {Azadi}}, \bibinfo
  {author} {\bibfnamefont {R.~V.}\ \bibnamefont {Belosludov}}, \bibinfo
  {author} {\bibfnamefont {T.~D.}\ \bibnamefont {K{\"u}hne}},\ and\ \bibinfo
  {author} {\bibfnamefont {M.~S.}\ \bibnamefont {Bahramy}},\ }\href
  {https://doi.org/10.48550/arXiv.2601.22065} {\bibinfo {title} {Universal
  {{Multifractality}} at the {{Topological Anderson Insulator Transition}}}}
  (\bibinfo {year} {2026}),\ \Eprint {https://arxiv.org/abs/2601.22065}
  {arXiv:2601.22065 [cond-mat.mtrl-sci]} \BibitemShut {NoStop}%
\bibitem [{\citenamefont {Biddle}\ and\ \citenamefont
  {Das~Sarma}(2010)}]{Biddle_Predicted_2010}%
  \BibitemOpen
  \bibfield  {author} {\bibinfo {author} {\bibfnamefont {J.}~\bibnamefont
  {Biddle}}\ and\ \bibinfo {author} {\bibfnamefont {S.}~\bibnamefont
  {Das~Sarma}},\ }\bibfield  {title} {\bibinfo {title} {Predicted {{Mobility
  Edges}} in {{One-Dimensional Incommensurate Optical Lattices}}: {{An Exactly
  Solvable Model}} of {{Anderson Localization}}},\ }\href
  {https://doi.org/10.1103/PhysRevLett.104.070601} {\bibfield  {journal}
  {\bibinfo  {journal} {Physical Review Letters}\ }\textbf {\bibinfo {volume}
  {104}},\ \bibinfo {pages} {070601} (\bibinfo {year} {2010})}\BibitemShut
  {NoStop}%
\bibitem [{\citenamefont {Biddle}\ \emph {et~al.}(2011)\citenamefont {Biddle},
  \citenamefont {Priour}, \citenamefont {Wang},\ and\ \citenamefont
  {Das~Sarma}}]{Biddle_Localization_2011}%
  \BibitemOpen
  \bibfield  {author} {\bibinfo {author} {\bibfnamefont {J.}~\bibnamefont
  {Biddle}}, \bibinfo {author} {\bibfnamefont {D.~J.}\ \bibnamefont {Priour}},
  \bibinfo {author} {\bibfnamefont {B.}~\bibnamefont {Wang}},\ and\ \bibinfo
  {author} {\bibfnamefont {S.}~\bibnamefont {Das~Sarma}},\ }\bibfield  {title}
  {\bibinfo {title} {Localization in one-dimensional lattices with
  non-nearest-neighbor hopping: {{Generalized Anderson}} and
  {{Aubry-Andr}}\textbackslash 'e models},\ }\href
  {https://doi.org/10.1103/PhysRevB.83.075105} {\bibfield  {journal} {\bibinfo
  {journal} {Physical Review B}\ }\textbf {\bibinfo {volume} {83}},\ \bibinfo
  {pages} {075105} (\bibinfo {year} {2011})}\BibitemShut {NoStop}%
\bibitem [{\citenamefont {Ganeshan}\ \emph {et~al.}(2015)\citenamefont
  {Ganeshan}, \citenamefont {Pixley},\ and\ \citenamefont
  {Das~Sarma}}]{Ganeshan_Nearest_2015}%
  \BibitemOpen
  \bibfield  {author} {\bibinfo {author} {\bibfnamefont {S.}~\bibnamefont
  {Ganeshan}}, \bibinfo {author} {\bibfnamefont {J.~H.}\ \bibnamefont
  {Pixley}},\ and\ \bibinfo {author} {\bibfnamefont {S.}~\bibnamefont
  {Das~Sarma}},\ }\bibfield  {title} {\bibinfo {title} {Nearest {{Neighbor
  Tight Binding Models}} with an {{Exact Mobility Edge}} in {{One
  Dimension}}},\ }\href {https://doi.org/10.1103/PhysRevLett.114.146601}
  {\bibfield  {journal} {\bibinfo  {journal} {Physical Review Letters}\
  }\textbf {\bibinfo {volume} {114}},\ \bibinfo {pages} {146601} (\bibinfo
  {year} {2015})}\BibitemShut {NoStop}%
\bibitem [{\citenamefont {Szab{\'o}}\ and\ \citenamefont
  {Schneider}(2020)}]{Szabo_Mixed_2020}%
  \BibitemOpen
  \bibfield  {author} {\bibinfo {author} {\bibfnamefont {A.}~\bibnamefont
  {Szab{\'o}}}\ and\ \bibinfo {author} {\bibfnamefont {U.}~\bibnamefont
  {Schneider}},\ }\bibfield  {title} {\bibinfo {title} {Mixed spectra and
  partially extended states in a two-dimensional quasiperiodic model},\ }\href
  {https://doi.org/10.1103/PhysRevB.101.014205} {\bibfield  {journal} {\bibinfo
   {journal} {Physical Review B}\ }\textbf {\bibinfo {volume} {101}},\ \bibinfo
  {pages} {014205} (\bibinfo {year} {2020})}\BibitemShut {NoStop}%
\bibitem [{\citenamefont {An}\ \emph {et~al.}(2021)\citenamefont {An},
  \citenamefont {Padavi{\'c}}, \citenamefont {Meier}, \citenamefont {Hegde},
  \citenamefont {Ganeshan}, \citenamefont {Pixley}, \citenamefont
  {Vishveshwara},\ and\ \citenamefont {Gadway}}]{An_Interactions_2021}%
  \BibitemOpen
  \bibfield  {author} {\bibinfo {author} {\bibfnamefont {F.~A.}\ \bibnamefont
  {An}}, \bibinfo {author} {\bibfnamefont {K.}~\bibnamefont {Padavi{\'c}}},
  \bibinfo {author} {\bibfnamefont {E.~J.}\ \bibnamefont {Meier}}, \bibinfo
  {author} {\bibfnamefont {S.}~\bibnamefont {Hegde}}, \bibinfo {author}
  {\bibfnamefont {S.}~\bibnamefont {Ganeshan}}, \bibinfo {author}
  {\bibfnamefont {J.~H.}\ \bibnamefont {Pixley}}, \bibinfo {author}
  {\bibfnamefont {S.}~\bibnamefont {Vishveshwara}},\ and\ \bibinfo {author}
  {\bibfnamefont {B.}~\bibnamefont {Gadway}},\ }\bibfield  {title} {\bibinfo
  {title} {Interactions and {{Mobility Edges}}: {{Observing}} the {{Generalized
  Aubry-Andr}}\textbackslash 'e {{Model}}},\ }\href
  {https://doi.org/10.1103/PhysRevLett.126.040603} {\bibfield  {journal}
  {\bibinfo  {journal} {Physical Review Letters}\ }\textbf {\bibinfo {volume}
  {126}},\ \bibinfo {pages} {040603} (\bibinfo {year} {2021})}\BibitemShut
  {NoStop}%
\bibitem [{\citenamefont {Zhu}\ \emph {et~al.}(2024)\citenamefont {Zhu},
  \citenamefont {Yu}, \citenamefont {Johnstone},\ and\ \citenamefont
  {{Sanchez-Palencia}}}]{Zhu_Localization_2024}%
  \BibitemOpen
  \bibfield  {author} {\bibinfo {author} {\bibfnamefont {Z.}~\bibnamefont
  {Zhu}}, \bibinfo {author} {\bibfnamefont {S.}~\bibnamefont {Yu}}, \bibinfo
  {author} {\bibfnamefont {D.}~\bibnamefont {Johnstone}},\ and\ \bibinfo
  {author} {\bibfnamefont {L.}~\bibnamefont {{Sanchez-Palencia}}},\ }\bibfield
  {title} {\bibinfo {title} {Localization and spectral structure in
  two-dimensional quasicrystal potentials},\ }\href
  {https://doi.org/10.1103/PhysRevA.109.013314} {\bibfield  {journal} {\bibinfo
   {journal} {Physical Review A}\ }\textbf {\bibinfo {volume} {109}},\ \bibinfo
  {pages} {013314} (\bibinfo {year} {2024})}\BibitemShut {NoStop}%
\bibitem [{\citenamefont {Duncan}(2024)}]{Duncan_Critical_2024}%
  \BibitemOpen
  \bibfield  {author} {\bibinfo {author} {\bibfnamefont {C.~W.}\ \bibnamefont
  {Duncan}},\ }\bibfield  {title} {\bibinfo {title} {Critical states and
  anomalous mobility edges in two-dimensional diagonal quasicrystals},\ }\href
  {https://doi.org/10.1103/PhysRevB.109.014210} {\bibfield  {journal} {\bibinfo
   {journal} {Physical Review B}\ }\textbf {\bibinfo {volume} {109}},\ \bibinfo
  {pages} {014210} (\bibinfo {year} {2024})}\BibitemShut {NoStop}%
\bibitem [{\citenamefont {Devakul}\ and\ \citenamefont
  {Huse}(2017)}]{Devakul_Anderson_2017}%
  \BibitemOpen
  \bibfield  {author} {\bibinfo {author} {\bibfnamefont {T.}~\bibnamefont
  {Devakul}}\ and\ \bibinfo {author} {\bibfnamefont {D.~A.}\ \bibnamefont
  {Huse}},\ }\bibfield  {title} {\bibinfo {title} {Anderson localization
  transitions with and without random potentials},\ }\href
  {https://doi.org/10.1103/PhysRevB.96.214201} {\bibfield  {journal} {\bibinfo
  {journal} {Physical Review B}\ }\textbf {\bibinfo {volume} {96}},\ \bibinfo
  {pages} {214201} (\bibinfo {year} {2017})}\BibitemShut {NoStop}%
\bibitem [{\citenamefont {Szab{\'o}}\ and\ \citenamefont
  {Schneider}(2018)}]{Szabo_Nonpowerlaw_2018}%
  \BibitemOpen
  \bibfield  {author} {\bibinfo {author} {\bibfnamefont {A.}~\bibnamefont
  {Szab{\'o}}}\ and\ \bibinfo {author} {\bibfnamefont {U.}~\bibnamefont
  {Schneider}},\ }\bibfield  {title} {\bibinfo {title} {Non-power-law
  universality in one-dimensional quasicrystals},\ }\href
  {https://doi.org/10.1103/PhysRevB.98.134201} {\bibfield  {journal} {\bibinfo
  {journal} {Physical Review B}\ }\textbf {\bibinfo {volume} {98}},\ \bibinfo
  {pages} {134201} (\bibinfo {year} {2018})}\BibitemShut {NoStop}%
\bibitem [{\citenamefont {{\v S}trkalj}\ \emph {et~al.}(2022)\citenamefont {{\v
  S}trkalj}, \citenamefont {Doggen},\ and\ \citenamefont
  {Castelnovo}}]{Strkalj_Coexistence_2022}%
  \BibitemOpen
  \bibfield  {author} {\bibinfo {author} {\bibfnamefont {A.}~\bibnamefont {{\v
  S}trkalj}}, \bibinfo {author} {\bibfnamefont {E.~V.~H.}\ \bibnamefont
  {Doggen}},\ and\ \bibinfo {author} {\bibfnamefont {C.}~\bibnamefont
  {Castelnovo}},\ }\bibfield  {title} {\bibinfo {title} {Coexistence of
  localization and transport in many-body two-dimensional
  {{Aubry-Andr}}\textbackslash 'e models},\ }\href
  {https://doi.org/10.1103/PhysRevB.106.184209} {\bibfield  {journal} {\bibinfo
   {journal} {Physical Review B}\ }\textbf {\bibinfo {volume} {106}},\ \bibinfo
  {pages} {184209} (\bibinfo {year} {2022})}\BibitemShut {NoStop}%
\bibitem [{\citenamefont {Thomson}\ and\ \citenamefont
  {Eisert}(2024)}]{Thomson_Unravelling_2024}%
  \BibitemOpen
  \bibfield  {author} {\bibinfo {author} {\bibfnamefont {S.~J.}\ \bibnamefont
  {Thomson}}\ and\ \bibinfo {author} {\bibfnamefont {J.}~\bibnamefont
  {Eisert}},\ }\bibfield  {title} {\bibinfo {title} {Unravelling quantum
  dynamics using flow equations},\ }\href
  {https://doi.org/10.1038/s41567-024-02549-2} {\bibfield  {journal} {\bibinfo
  {journal} {Nature Physics}\ }\textbf {\bibinfo {volume} {20}},\ \bibinfo
  {pages} {1401} (\bibinfo {year} {2024})}\BibitemShut {NoStop}%
\bibitem [{\citenamefont {Sbroscia}\ \emph {et~al.}(2020)\citenamefont
  {Sbroscia}, \citenamefont {Viebahn}, \citenamefont {Carter}, \citenamefont
  {Yu}, \citenamefont {Gaunt},\ and\ \citenamefont
  {Schneider}}]{Sbroscia_Observing_2020}%
  \BibitemOpen
  \bibfield  {author} {\bibinfo {author} {\bibfnamefont {M.}~\bibnamefont
  {Sbroscia}}, \bibinfo {author} {\bibfnamefont {K.}~\bibnamefont {Viebahn}},
  \bibinfo {author} {\bibfnamefont {E.}~\bibnamefont {Carter}}, \bibinfo
  {author} {\bibfnamefont {J.-C.}\ \bibnamefont {Yu}}, \bibinfo {author}
  {\bibfnamefont {A.}~\bibnamefont {Gaunt}},\ and\ \bibinfo {author}
  {\bibfnamefont {U.}~\bibnamefont {Schneider}},\ }\bibfield  {title} {\bibinfo
  {title} {Observing {{Localization}} in a {{2D Quasicrystalline Optical
  Lattice}}},\ }\href {https://doi.org/10.1103/PhysRevLett.125.200604}
  {\bibfield  {journal} {\bibinfo  {journal} {Physical Review Letters}\
  }\textbf {\bibinfo {volume} {125}},\ \bibinfo {pages} {200604} (\bibinfo
  {year} {2020})}\BibitemShut {NoStop}%
\bibitem [{\citenamefont {Hur}\ \emph {et~al.}(2025)\citenamefont {Hur},
  \citenamefont {Li}, \citenamefont {Lee}, \citenamefont {Kwon}, \citenamefont
  {Kim}, \citenamefont {Hwang}, \citenamefont {Kim}, \citenamefont {Yu},
  \citenamefont {Chan}, \citenamefont {Wahl},\ and\ \citenamefont
  {Choi}}]{Hur_Stability_2025}%
  \BibitemOpen
  \bibfield  {author} {\bibinfo {author} {\bibfnamefont {J.}~\bibnamefont
  {Hur}}, \bibinfo {author} {\bibfnamefont {J.}~\bibnamefont {Li}}, \bibinfo
  {author} {\bibfnamefont {B.}~\bibnamefont {Lee}}, \bibinfo {author}
  {\bibfnamefont {K.}~\bibnamefont {Kwon}}, \bibinfo {author} {\bibfnamefont
  {M.}~\bibnamefont {Kim}}, \bibinfo {author} {\bibfnamefont {S.}~\bibnamefont
  {Hwang}}, \bibinfo {author} {\bibfnamefont {S.}~\bibnamefont {Kim}}, \bibinfo
  {author} {\bibfnamefont {Y.~S.}\ \bibnamefont {Yu}}, \bibinfo {author}
  {\bibfnamefont {A.}~\bibnamefont {Chan}}, \bibinfo {author} {\bibfnamefont
  {T.}~\bibnamefont {Wahl}},\ and\ \bibinfo {author} {\bibfnamefont {J.-y.}\
  \bibnamefont {Choi}},\ }\href {https://doi.org/10.48550/arXiv.2508.20699}
  {\bibinfo {title} {Stability of many-body localization in two dimensions}}
  (\bibinfo {year} {2025}),\ \Eprint {https://arxiv.org/abs/2508.20699}
  {arXiv:2508.20699 [cond-mat]} \BibitemShut {NoStop}%
\bibitem [{\citenamefont {De~Roeck}\ and\ \citenamefont
  {Huveneers}(2017)}]{DeRoeck_Stability_2017}%
  \BibitemOpen
  \bibfield  {author} {\bibinfo {author} {\bibfnamefont {W.}~\bibnamefont
  {De~Roeck}}\ and\ \bibinfo {author} {\bibfnamefont {F.}~\bibnamefont
  {Huveneers}},\ }\bibfield  {title} {\bibinfo {title} {Stability and
  instability towards delocalization in many-body localization systems},\
  }\href {https://doi.org/10.1103/PhysRevB.95.155129} {\bibfield  {journal}
  {\bibinfo  {journal} {Physical Review B}\ }\textbf {\bibinfo {volume} {95}},\
  \bibinfo {pages} {155129} (\bibinfo {year} {2017})}\BibitemShut {NoStop}%
\bibitem [{\citenamefont {Luitz}\ \emph {et~al.}(2017)\citenamefont {Luitz},
  \citenamefont {Huveneers},\ and\ \citenamefont {De~Roeck}}]{Luitz_How_2017}%
  \BibitemOpen
  \bibfield  {author} {\bibinfo {author} {\bibfnamefont {D.~J.}\ \bibnamefont
  {Luitz}}, \bibinfo {author} {\bibfnamefont {F.}~\bibnamefont {Huveneers}},\
  and\ \bibinfo {author} {\bibfnamefont {W.}~\bibnamefont {De~Roeck}},\
  }\bibfield  {title} {\bibinfo {title} {How a {{Small Quantum Bath Can
  Thermalize Long Localized Chains}}},\ }\href
  {https://doi.org/10.1103/PhysRevLett.119.150602} {\bibfield  {journal}
  {\bibinfo  {journal} {Physical Review Letters}\ }\textbf {\bibinfo {volume}
  {119}},\ \bibinfo {pages} {150602} (\bibinfo {year} {2017})}\BibitemShut
  {NoStop}%
\bibitem [{\citenamefont {Thiery}\ \emph {et~al.}(2018)\citenamefont {Thiery},
  \citenamefont {Huveneers}, \citenamefont {M{\"u}ller},\ and\ \citenamefont
  {De~Roeck}}]{Thiery_ManyBody_2018}%
  \BibitemOpen
  \bibfield  {author} {\bibinfo {author} {\bibfnamefont {T.}~\bibnamefont
  {Thiery}}, \bibinfo {author} {\bibfnamefont {F.}~\bibnamefont {Huveneers}},
  \bibinfo {author} {\bibfnamefont {M.}~\bibnamefont {M{\"u}ller}},\ and\
  \bibinfo {author} {\bibfnamefont {W.}~\bibnamefont {De~Roeck}},\ }\bibfield
  {title} {\bibinfo {title} {Many-{{Body Delocalization}} as a {{Quantum
  Avalanche}}},\ }\href {https://doi.org/10.1103/PhysRevLett.121.140601}
  {\bibfield  {journal} {\bibinfo  {journal} {Physical Review Letters}\
  }\textbf {\bibinfo {volume} {121}},\ \bibinfo {pages} {140601} (\bibinfo
  {year} {2018})}\BibitemShut {NoStop}%
\bibitem [{\citenamefont {Morningstar}\ \emph {et~al.}(2022)\citenamefont
  {Morningstar}, \citenamefont {Colmenarez}, \citenamefont {Khemani},
  \citenamefont {Luitz},\ and\ \citenamefont
  {Huse}}]{Morningstar_Avalanches_2022}%
  \BibitemOpen
  \bibfield  {author} {\bibinfo {author} {\bibfnamefont {A.}~\bibnamefont
  {Morningstar}}, \bibinfo {author} {\bibfnamefont {L.}~\bibnamefont
  {Colmenarez}}, \bibinfo {author} {\bibfnamefont {V.}~\bibnamefont {Khemani}},
  \bibinfo {author} {\bibfnamefont {D.~J.}\ \bibnamefont {Luitz}},\ and\
  \bibinfo {author} {\bibfnamefont {D.~A.}\ \bibnamefont {Huse}},\ }\bibfield
  {title} {\bibinfo {title} {Avalanches and many-body resonances in many-body
  localized systems},\ }\href {https://doi.org/10.1103/PhysRevB.105.174205}
  {\bibfield  {journal} {\bibinfo  {journal} {Physical Review B}\ }\textbf
  {\bibinfo {volume} {105}},\ \bibinfo {pages} {174205} (\bibinfo {year}
  {2022})}\BibitemShut {NoStop}%
\bibitem [{\citenamefont {Ahn}\ \emph {et~al.}(2018)\citenamefont {Ahn},
  \citenamefont {Moon}, \citenamefont {Kim}, \citenamefont {Kim}, \citenamefont
  {Shin}, \citenamefont {Kim}, \citenamefont {Cha}, \citenamefont {Kahng},
  \citenamefont {Kim}, \citenamefont {Koshino}, \citenamefont {Son},
  \citenamefont {Yang},\ and\ \citenamefont {Ahn}}]{Ahn_Dirac_2018}%
  \BibitemOpen
  \bibfield  {author} {\bibinfo {author} {\bibfnamefont {S.~J.}\ \bibnamefont
  {Ahn}}, \bibinfo {author} {\bibfnamefont {P.}~\bibnamefont {Moon}}, \bibinfo
  {author} {\bibfnamefont {T.-H.}\ \bibnamefont {Kim}}, \bibinfo {author}
  {\bibfnamefont {H.-W.}\ \bibnamefont {Kim}}, \bibinfo {author} {\bibfnamefont
  {H.-C.}\ \bibnamefont {Shin}}, \bibinfo {author} {\bibfnamefont {E.~H.}\
  \bibnamefont {Kim}}, \bibinfo {author} {\bibfnamefont {H.~W.}\ \bibnamefont
  {Cha}}, \bibinfo {author} {\bibfnamefont {S.-J.}\ \bibnamefont {Kahng}},
  \bibinfo {author} {\bibfnamefont {P.}~\bibnamefont {Kim}}, \bibinfo {author}
  {\bibfnamefont {M.}~\bibnamefont {Koshino}}, \bibinfo {author} {\bibfnamefont
  {Y.-W.}\ \bibnamefont {Son}}, \bibinfo {author} {\bibfnamefont {C.-W.}\
  \bibnamefont {Yang}},\ and\ \bibinfo {author} {\bibfnamefont {J.~R.}\
  \bibnamefont {Ahn}},\ }\bibfield  {title} {\bibinfo {title} {Dirac electrons
  in a dodecagonal graphene quasicrystal},\ }\href
  {https://doi.org/10.1126/science.aar8412} {\bibfield  {journal} {\bibinfo
  {journal} {Science}\ }\textbf {\bibinfo {volume} {361}},\ \bibinfo {pages}
  {782} (\bibinfo {year} {2018})}\BibitemShut {NoStop}%
\bibitem [{\citenamefont {Yao}\ \emph {et~al.}(2018)\citenamefont {Yao},
  \citenamefont {Wang}, \citenamefont {Bao}, \citenamefont {Zhang},
  \citenamefont {Zhang}, \citenamefont {Bao}, \citenamefont {Chan},
  \citenamefont {Chen}, \citenamefont {Avila}, \citenamefont {Asensio},
  \citenamefont {Zhu},\ and\ \citenamefont {Zhou}}]{Yao_Quasicrystalline_2018}%
  \BibitemOpen
  \bibfield  {author} {\bibinfo {author} {\bibfnamefont {W.}~\bibnamefont
  {Yao}}, \bibinfo {author} {\bibfnamefont {E.}~\bibnamefont {Wang}}, \bibinfo
  {author} {\bibfnamefont {C.}~\bibnamefont {Bao}}, \bibinfo {author}
  {\bibfnamefont {Y.}~\bibnamefont {Zhang}}, \bibinfo {author} {\bibfnamefont
  {K.}~\bibnamefont {Zhang}}, \bibinfo {author} {\bibfnamefont
  {K.}~\bibnamefont {Bao}}, \bibinfo {author} {\bibfnamefont {C.~K.}\
  \bibnamefont {Chan}}, \bibinfo {author} {\bibfnamefont {C.}~\bibnamefont
  {Chen}}, \bibinfo {author} {\bibfnamefont {J.}~\bibnamefont {Avila}},
  \bibinfo {author} {\bibfnamefont {M.~C.}\ \bibnamefont {Asensio}}, \bibinfo
  {author} {\bibfnamefont {J.}~\bibnamefont {Zhu}},\ and\ \bibinfo {author}
  {\bibfnamefont {S.}~\bibnamefont {Zhou}},\ }\bibfield  {title} {\bibinfo
  {title} {Quasicrystalline 30{$^\circ$} twisted bilayer graphene as an
  incommensurate superlattice with strong interlayer coupling},\ }\href
  {https://doi.org/10.1073/pnas.1720865115} {\bibfield  {journal} {\bibinfo
  {journal} {Proceedings of the National Academy of Sciences}\ }\textbf
  {\bibinfo {volume} {115}},\ \bibinfo {pages} {6928} (\bibinfo {year}
  {2018})}\BibitemShut {NoStop}%
\bibitem [{\citenamefont {Yu}\ \emph {et~al.}(2019)\citenamefont {Yu},
  \citenamefont {Wu}, \citenamefont {Zhan}, \citenamefont {Katsnelson},\ and\
  \citenamefont {Yuan}}]{Yu_Dodecagonal_2019}%
  \BibitemOpen
  \bibfield  {author} {\bibinfo {author} {\bibfnamefont {G.}~\bibnamefont
  {Yu}}, \bibinfo {author} {\bibfnamefont {Z.}~\bibnamefont {Wu}}, \bibinfo
  {author} {\bibfnamefont {Z.}~\bibnamefont {Zhan}}, \bibinfo {author}
  {\bibfnamefont {M.~I.}\ \bibnamefont {Katsnelson}},\ and\ \bibinfo {author}
  {\bibfnamefont {S.}~\bibnamefont {Yuan}},\ }\bibfield  {title} {\bibinfo
  {title} {Dodecagonal bilayer graphene quasicrystal and its approximants},\
  }\href {https://doi.org/10.1038/s41524-019-0258-0} {\bibfield  {journal}
  {\bibinfo  {journal} {npj Computational Materials}\ }\textbf {\bibinfo
  {volume} {5}},\ \bibinfo {pages} {1} (\bibinfo {year} {2019})}\BibitemShut
  {NoStop}%
\bibitem [{\citenamefont {Moon}\ \emph {et~al.}(2019)\citenamefont {Moon},
  \citenamefont {Koshino},\ and\ \citenamefont
  {Son}}]{Moon_Quasicrystalline_2019}%
  \BibitemOpen
  \bibfield  {author} {\bibinfo {author} {\bibfnamefont {P.}~\bibnamefont
  {Moon}}, \bibinfo {author} {\bibfnamefont {M.}~\bibnamefont {Koshino}},\ and\
  \bibinfo {author} {\bibfnamefont {Y.-W.}\ \bibnamefont {Son}},\ }\bibfield
  {title} {\bibinfo {title} {Quasicrystalline electronic states in 30 {$\circ$}
  rotated twisted bilayer graphene},\ }\href
  {https://doi.org/10.1103/PhysRevB.99.165430} {\bibfield  {journal} {\bibinfo
  {journal} {Physical Review B}\ }\textbf {\bibinfo {volume} {99}},\ \bibinfo
  {pages} {165430} (\bibinfo {year} {2019})}\BibitemShut {NoStop}%
\bibitem [{\citenamefont {Crosse}\ and\ \citenamefont
  {Moon}(2021)}]{Crosse_Quasicrystalline_2021}%
  \BibitemOpen
  \bibfield  {author} {\bibinfo {author} {\bibfnamefont {J.~A.}\ \bibnamefont
  {Crosse}}\ and\ \bibinfo {author} {\bibfnamefont {P.}~\bibnamefont {Moon}},\
  }\bibfield  {title} {\bibinfo {title} {Quasicrystalline electronic states in
  twisted bilayers and the effects of interlayer and sublattice symmetries},\
  }\href {https://doi.org/10.1103/PhysRevB.103.045408} {\bibfield  {journal}
  {\bibinfo  {journal} {Physical Review B}\ }\textbf {\bibinfo {volume}
  {103}},\ \bibinfo {pages} {045408} (\bibinfo {year} {2021})}\BibitemShut
  {NoStop}%
\bibitem [{\citenamefont {Mao}\ and\ \citenamefont
  {Senthil}(2021)}]{Mao_Quasiperiodicity_2021}%
  \BibitemOpen
  \bibfield  {author} {\bibinfo {author} {\bibfnamefont {D.}~\bibnamefont
  {Mao}}\ and\ \bibinfo {author} {\bibfnamefont {T.}~\bibnamefont {Senthil}},\
  }\bibfield  {title} {\bibinfo {title} {Quasiperiodicity, band topology, and
  moir\'e graphene},\ }\href {https://doi.org/10.1103/PhysRevB.103.115110}
  {\bibfield  {journal} {\bibinfo  {journal} {Physical Review B}\ }\textbf
  {\bibinfo {volume} {103}},\ \bibinfo {pages} {115110} (\bibinfo {year}
  {2021})}\BibitemShut {NoStop}%
\bibitem [{\citenamefont {Uri}\ \emph {et~al.}(2023)\citenamefont {Uri},
  \citenamefont {{de la Barrera}}, \citenamefont {Randeria}, \citenamefont
  {{Rodan-Legrain}}, \citenamefont {Devakul}, \citenamefont {Crowley},
  \citenamefont {Paul}, \citenamefont {Watanabe}, \citenamefont {Taniguchi},
  \citenamefont {Lifshitz}, \citenamefont {Fu}, \citenamefont {Ashoori},\ and\
  \citenamefont {{Jarillo-Herrero}}}]{Uri_Superconductivity_2023}%
  \BibitemOpen
  \bibfield  {author} {\bibinfo {author} {\bibfnamefont {A.}~\bibnamefont
  {Uri}}, \bibinfo {author} {\bibfnamefont {S.~C.}\ \bibnamefont {{de la
  Barrera}}}, \bibinfo {author} {\bibfnamefont {M.~T.}\ \bibnamefont
  {Randeria}}, \bibinfo {author} {\bibfnamefont {D.}~\bibnamefont
  {{Rodan-Legrain}}}, \bibinfo {author} {\bibfnamefont {T.}~\bibnamefont
  {Devakul}}, \bibinfo {author} {\bibfnamefont {P.~J.~D.}\ \bibnamefont
  {Crowley}}, \bibinfo {author} {\bibfnamefont {N.}~\bibnamefont {Paul}},
  \bibinfo {author} {\bibfnamefont {K.}~\bibnamefont {Watanabe}}, \bibinfo
  {author} {\bibfnamefont {T.}~\bibnamefont {Taniguchi}}, \bibinfo {author}
  {\bibfnamefont {R.}~\bibnamefont {Lifshitz}}, \bibinfo {author}
  {\bibfnamefont {L.}~\bibnamefont {Fu}}, \bibinfo {author} {\bibfnamefont
  {R.~C.}\ \bibnamefont {Ashoori}},\ and\ \bibinfo {author} {\bibfnamefont
  {P.}~\bibnamefont {{Jarillo-Herrero}}},\ }\bibfield  {title} {\bibinfo
  {title} {Superconductivity and strong interactions in a tunable moir\'e
  quasicrystal},\ }\href {https://doi.org/10.1038/s41586-023-06294-z}
  {\bibfield  {journal} {\bibinfo  {journal} {Nature}\ }\textbf {\bibinfo
  {volume} {620}},\ \bibinfo {pages} {762} (\bibinfo {year}
  {2023})}\BibitemShut {NoStop}%
\bibitem [{\citenamefont {Roati}\ \emph {et~al.}(2008)\citenamefont {Roati},
  \citenamefont {D'Errico}, \citenamefont {Fallani}, \citenamefont {Fattori},
  \citenamefont {Fort}, \citenamefont {Zaccanti}, \citenamefont {Modugno},
  \citenamefont {Modugno},\ and\ \citenamefont
  {Inguscio}}]{Roati_Anderson_2008}%
  \BibitemOpen
  \bibfield  {author} {\bibinfo {author} {\bibfnamefont {G.}~\bibnamefont
  {Roati}}, \bibinfo {author} {\bibfnamefont {C.}~\bibnamefont {D'Errico}},
  \bibinfo {author} {\bibfnamefont {L.}~\bibnamefont {Fallani}}, \bibinfo
  {author} {\bibfnamefont {M.}~\bibnamefont {Fattori}}, \bibinfo {author}
  {\bibfnamefont {C.}~\bibnamefont {Fort}}, \bibinfo {author} {\bibfnamefont
  {M.}~\bibnamefont {Zaccanti}}, \bibinfo {author} {\bibfnamefont
  {G.}~\bibnamefont {Modugno}}, \bibinfo {author} {\bibfnamefont
  {M.}~\bibnamefont {Modugno}},\ and\ \bibinfo {author} {\bibfnamefont
  {M.}~\bibnamefont {Inguscio}},\ }\bibfield  {title} {\bibinfo {title}
  {Anderson localization of a non-interacting {{Bose}}--{{Einstein}}
  condensate},\ }\href {https://doi.org/10.1038/nature07071} {\bibfield
  {journal} {\bibinfo  {journal} {Nature}\ }\textbf {\bibinfo {volume} {453}},\
  \bibinfo {pages} {895} (\bibinfo {year} {2008})}\BibitemShut {NoStop}%
\bibitem [{\citenamefont {Schreiber}\ \emph {et~al.}(2015)\citenamefont
  {Schreiber}, \citenamefont {Hodgman}, \citenamefont {Bordia}, \citenamefont
  {L{\"u}schen}, \citenamefont {Fischer}, \citenamefont {Vosk}, \citenamefont
  {Altman}, \citenamefont {Schneider},\ and\ \citenamefont
  {Bloch}}]{Schreiber_Observation_2015}%
  \BibitemOpen
  \bibfield  {author} {\bibinfo {author} {\bibfnamefont {M.}~\bibnamefont
  {Schreiber}}, \bibinfo {author} {\bibfnamefont {S.~S.}\ \bibnamefont
  {Hodgman}}, \bibinfo {author} {\bibfnamefont {P.}~\bibnamefont {Bordia}},
  \bibinfo {author} {\bibfnamefont {H.~P.}\ \bibnamefont {L{\"u}schen}},
  \bibinfo {author} {\bibfnamefont {M.~H.}\ \bibnamefont {Fischer}}, \bibinfo
  {author} {\bibfnamefont {R.}~\bibnamefont {Vosk}}, \bibinfo {author}
  {\bibfnamefont {E.}~\bibnamefont {Altman}}, \bibinfo {author} {\bibfnamefont
  {U.}~\bibnamefont {Schneider}},\ and\ \bibinfo {author} {\bibfnamefont
  {I.}~\bibnamefont {Bloch}},\ }\bibfield  {title} {\bibinfo {title}
  {Observation of many-body localization of interacting fermions in a
  quasirandom optical lattice},\ }\href
  {https://doi.org/10.1126/science.aaa7432} {\bibfield  {journal} {\bibinfo
  {journal} {Science}\ }\textbf {\bibinfo {volume} {349}},\ \bibinfo {pages}
  {842} (\bibinfo {year} {2015})}\BibitemShut {NoStop}%
\bibitem [{\citenamefont {L{\"u}schen}\ \emph {et~al.}(2017)\citenamefont
  {L{\"u}schen}, \citenamefont {Bordia}, \citenamefont {Scherg}, \citenamefont
  {Alet}, \citenamefont {Altman}, \citenamefont {Schneider},\ and\
  \citenamefont {Bloch}}]{Luschen_Observation_2017}%
  \BibitemOpen
  \bibfield  {author} {\bibinfo {author} {\bibfnamefont {H.~P.}\ \bibnamefont
  {L{\"u}schen}}, \bibinfo {author} {\bibfnamefont {P.}~\bibnamefont {Bordia}},
  \bibinfo {author} {\bibfnamefont {S.}~\bibnamefont {Scherg}}, \bibinfo
  {author} {\bibfnamefont {F.}~\bibnamefont {Alet}}, \bibinfo {author}
  {\bibfnamefont {E.}~\bibnamefont {Altman}}, \bibinfo {author} {\bibfnamefont
  {U.}~\bibnamefont {Schneider}},\ and\ \bibinfo {author} {\bibfnamefont
  {I.}~\bibnamefont {Bloch}},\ }\bibfield  {title} {\bibinfo {title}
  {Observation of {{Slow Dynamics}} near the {{Many-Body Localization
  Transition}} in {{One-Dimensional Quasiperiodic Systems}}},\ }\href
  {https://doi.org/10.1103/PhysRevLett.119.260401} {\bibfield  {journal}
  {\bibinfo  {journal} {Physical Review Letters}\ }\textbf {\bibinfo {volume}
  {119}},\ \bibinfo {pages} {260401} (\bibinfo {year} {2017})}\BibitemShut
  {NoStop}%
\bibitem [{\citenamefont {Bordia}\ \emph {et~al.}(2016)\citenamefont {Bordia},
  \citenamefont {L{\"u}schen}, \citenamefont {Hodgman}, \citenamefont
  {Schreiber}, \citenamefont {Bloch},\ and\ \citenamefont
  {Schneider}}]{Bordia_Coupling_2016}%
  \BibitemOpen
  \bibfield  {author} {\bibinfo {author} {\bibfnamefont {P.}~\bibnamefont
  {Bordia}}, \bibinfo {author} {\bibfnamefont {H.~P.}\ \bibnamefont
  {L{\"u}schen}}, \bibinfo {author} {\bibfnamefont {S.~S.}\ \bibnamefont
  {Hodgman}}, \bibinfo {author} {\bibfnamefont {M.}~\bibnamefont {Schreiber}},
  \bibinfo {author} {\bibfnamefont {I.}~\bibnamefont {Bloch}},\ and\ \bibinfo
  {author} {\bibfnamefont {U.}~\bibnamefont {Schneider}},\ }\bibfield  {title}
  {\bibinfo {title} {Coupling {{Identical}} one-dimensional {{Many-Body
  Localized Systems}}},\ }\href
  {https://doi.org/10.1103/PhysRevLett.116.140401} {\bibfield  {journal}
  {\bibinfo  {journal} {Physical Review Letters}\ }\textbf {\bibinfo {volume}
  {116}},\ \bibinfo {pages} {140401} (\bibinfo {year} {2016})}\BibitemShut
  {NoStop}%
\bibitem [{\citenamefont {Bordia}\ \emph {et~al.}(2017)\citenamefont {Bordia},
  \citenamefont {L{\"u}schen}, \citenamefont {Scherg}, \citenamefont
  {Gopalakrishnan}, \citenamefont {Knap}, \citenamefont {Schneider},\ and\
  \citenamefont {Bloch}}]{Bordia_Probing_2017}%
  \BibitemOpen
  \bibfield  {author} {\bibinfo {author} {\bibfnamefont {P.}~\bibnamefont
  {Bordia}}, \bibinfo {author} {\bibfnamefont {H.}~\bibnamefont {L{\"u}schen}},
  \bibinfo {author} {\bibfnamefont {S.}~\bibnamefont {Scherg}}, \bibinfo
  {author} {\bibfnamefont {S.}~\bibnamefont {Gopalakrishnan}}, \bibinfo
  {author} {\bibfnamefont {M.}~\bibnamefont {Knap}}, \bibinfo {author}
  {\bibfnamefont {U.}~\bibnamefont {Schneider}},\ and\ \bibinfo {author}
  {\bibfnamefont {I.}~\bibnamefont {Bloch}},\ }\bibfield  {title} {\bibinfo
  {title} {Probing {{Slow Relaxation}} and {{Many-Body Localization}} in
  {{Two-Dimensional Quasiperiodic Systems}}},\ }\href
  {https://doi.org/10.1103/PhysRevX.7.041047} {\bibfield  {journal} {\bibinfo
  {journal} {Physical Review X}\ }\textbf {\bibinfo {volume} {7}},\ \bibinfo
  {pages} {041047} (\bibinfo {year} {2017})}\BibitemShut {NoStop}%
\bibitem [{\citenamefont {Viebahn}\ \emph {et~al.}(2019)\citenamefont
  {Viebahn}, \citenamefont {Sbroscia}, \citenamefont {Carter}, \citenamefont
  {Yu},\ and\ \citenamefont {Schneider}}]{Viebahn_MatterWave_2019}%
  \BibitemOpen
  \bibfield  {author} {\bibinfo {author} {\bibfnamefont {K.}~\bibnamefont
  {Viebahn}}, \bibinfo {author} {\bibfnamefont {M.}~\bibnamefont {Sbroscia}},
  \bibinfo {author} {\bibfnamefont {E.}~\bibnamefont {Carter}}, \bibinfo
  {author} {\bibfnamefont {J.-C.}\ \bibnamefont {Yu}},\ and\ \bibinfo {author}
  {\bibfnamefont {U.}~\bibnamefont {Schneider}},\ }\bibfield  {title} {\bibinfo
  {title} {Matter-{{Wave Diffraction}} from a {{Quasicrystalline Optical
  Lattice}}},\ }\href {https://doi.org/10.1103/PhysRevLett.122.110404}
  {\bibfield  {journal} {\bibinfo  {journal} {Physical Review Letters}\
  }\textbf {\bibinfo {volume} {122}},\ \bibinfo {pages} {110404} (\bibinfo
  {year} {2019})}\BibitemShut {NoStop}%
\bibitem [{\citenamefont {Burgess}\ and\ \citenamefont
  {Cooper}(2026)}]{Burgess_Quasicrystalline_2026}%
  \BibitemOpen
  \bibfield  {author} {\bibinfo {author} {\bibfnamefont {B.~D.~K.}\
  \bibnamefont {Burgess}}\ and\ \bibinfo {author} {\bibfnamefont {N.~R.}\
  \bibnamefont {Cooper}},\ }\bibfield  {title} {\bibinfo {title}
  {Quasicrystalline analog of the {{Haldane}} model},\ }\href
  {https://doi.org/10.1103/2556-3lw6} {\bibfield  {journal} {\bibinfo
  {journal} {Physical Review Research}\ }\textbf {\bibinfo {volume} {8}},\
  \bibinfo {pages} {033105} (\bibinfo {year} {2026})}\BibitemShut {NoStop}%
\bibitem [{\citenamefont {Neely}\ \emph {et~al.}(2026)\citenamefont {Neely},
  \citenamefont {Wilson}, \citenamefont {Everly}, \citenamefont {Yao},
  \citenamefont {Zanetti},\ and\ \citenamefont
  {Brown}}]{Neely_Programmable_2026}%
  \BibitemOpen
  \bibfield  {author} {\bibinfo {author} {\bibfnamefont {A.~O.}\ \bibnamefont
  {Neely}}, \bibinfo {author} {\bibfnamefont {C.~C.}\ \bibnamefont {Wilson}},
  \bibinfo {author} {\bibfnamefont {R.}~\bibnamefont {Everly}}, \bibinfo
  {author} {\bibfnamefont {Y.}~\bibnamefont {Yao}}, \bibinfo {author}
  {\bibfnamefont {R.~F.}\ \bibnamefont {Zanetti}},\ and\ \bibinfo {author}
  {\bibfnamefont {C.~D.}\ \bibnamefont {Brown}},\ }\bibfield  {title} {\bibinfo
  {title} {Programmable dynamic phase control of a quasiperiodic optical
  lattice},\ }\href {https://doi.org/10.1103/z5xd-vqh9} {\bibfield  {journal}
  {\bibinfo  {journal} {Physical Review Applied}\ }\textbf {\bibinfo {volume}
  {26}},\ \bibinfo {pages} {034016} (\bibinfo {year} {2026})}\BibitemShut
  {NoStop}%
\bibitem [{\citenamefont {Fu}\ \emph {et~al.}(2021)\citenamefont {Fu},
  \citenamefont {Wilson},\ and\ \citenamefont {Pixley}}]{Fu_Flat_2021}%
  \BibitemOpen
  \bibfield  {author} {\bibinfo {author} {\bibfnamefont {Y.}~\bibnamefont
  {Fu}}, \bibinfo {author} {\bibfnamefont {J.~H.}\ \bibnamefont {Wilson}},\
  and\ \bibinfo {author} {\bibfnamefont {J.~H.}\ \bibnamefont {Pixley}},\
  }\bibfield  {title} {\bibinfo {title} {Flat topological bands and eigenstate
  criticality in a quasiperiodic insulator},\ }\href
  {https://doi.org/10.1103/PhysRevB.104.L041106} {\bibfield  {journal}
  {\bibinfo  {journal} {Physical Review B}\ }\textbf {\bibinfo {volume}
  {104}},\ \bibinfo {pages} {L041106} (\bibinfo {year} {2021})}\BibitemShut
  {NoStop}%
\bibitem [{\citenamefont {Madeira}\ and\ \citenamefont
  {Sacramento}(2022)}]{Madeira_Quasidisorderinduced_2022}%
  \BibitemOpen
  \bibfield  {author} {\bibinfo {author} {\bibfnamefont {M.~F.}\ \bibnamefont
  {Madeira}}\ and\ \bibinfo {author} {\bibfnamefont {P.~D.}\ \bibnamefont
  {Sacramento}},\ }\bibfield  {title} {\bibinfo {title} {Quasidisorder-induced
  topology},\ }\href {https://doi.org/10.1103/PhysRevB.106.224505} {\bibfield
  {journal} {\bibinfo  {journal} {Physical Review B}\ }\textbf {\bibinfo
  {volume} {106}},\ \bibinfo {pages} {224505} (\bibinfo {year}
  {2022})}\BibitemShut {NoStop}%
\bibitem [{\citenamefont {Gon{\c c}alves}\ \emph {et~al.}(2022)\citenamefont
  {Gon{\c c}alves}, \citenamefont {Gon{\c c}alves}, \citenamefont {Ribeiro},\
  and\ \citenamefont {Amorim}}]{Goncalves_Topological_2022}%
  \BibitemOpen
  \bibfield  {author} {\bibinfo {author} {\bibfnamefont {T.~S.}\ \bibnamefont
  {Gon{\c c}alves}}, \bibinfo {author} {\bibfnamefont {M.}~\bibnamefont {Gon{\c
  c}alves}}, \bibinfo {author} {\bibfnamefont {P.}~\bibnamefont {Ribeiro}},\
  and\ \bibinfo {author} {\bibfnamefont {B.}~\bibnamefont {Amorim}},\ }\href
  {https://doi.org/10.48550/arXiv.2212.08024} {\bibinfo {title} {Topological
  phase transitions for any taste in {{2D}} quasiperiodic systems}} (\bibinfo
  {year} {2022}),\ \Eprint {https://arxiv.org/abs/2212.08024} {arXiv:2212.08024
  [cond-mat.dis-nn]} \BibitemShut {NoStop}%
\bibitem [{\citenamefont {Cheng}\ \emph {et~al.}(2023)\citenamefont {Cheng},
  \citenamefont {Asgari},\ and\ \citenamefont
  {Xianlong}}]{Cheng_Topological_2023}%
  \BibitemOpen
  \bibfield  {author} {\bibinfo {author} {\bibfnamefont {S.}~\bibnamefont
  {Cheng}}, \bibinfo {author} {\bibfnamefont {R.}~\bibnamefont {Asgari}},\ and\
  \bibinfo {author} {\bibfnamefont {G.}~\bibnamefont {Xianlong}},\ }\bibfield
  {title} {\bibinfo {title} {From topological phase to transverse {{Anderson}}
  localization in a two-dimensional quasiperiodic system},\ }\href
  {https://doi.org/10.1103/PhysRevB.108.024204} {\bibfield  {journal} {\bibinfo
   {journal} {Physical Review B}\ }\textbf {\bibinfo {volume} {108}},\ \bibinfo
  {pages} {024204} (\bibinfo {year} {2023})}\BibitemShut {NoStop}%
\bibitem [{\citenamefont {Liquito}\ \emph {et~al.}(2024)\citenamefont
  {Liquito}, \citenamefont {Gon{\c c}alves},\ and\ \citenamefont
  {Castro}}]{Liquito_Fate_2024}%
  \BibitemOpen
  \bibfield  {author} {\bibinfo {author} {\bibfnamefont {R.}~\bibnamefont
  {Liquito}}, \bibinfo {author} {\bibfnamefont {M.}~\bibnamefont {Gon{\c
  c}alves}},\ and\ \bibinfo {author} {\bibfnamefont {E.~V.}\ \bibnamefont
  {Castro}},\ }\bibfield  {title} {\bibinfo {title} {Fate of quadratic band
  crossing under quasiperiodic modulation},\ }\href
  {https://doi.org/10.1103/PhysRevB.109.174202} {\bibfield  {journal} {\bibinfo
   {journal} {Physical Review B}\ }\textbf {\bibinfo {volume} {109}},\ \bibinfo
  {pages} {174202} (\bibinfo {year} {2024})}\BibitemShut {NoStop}%
\bibitem [{\citenamefont {Karcher}\ \emph {et~al.}(2024)\citenamefont
  {Karcher}, \citenamefont {Vasseur},\ and\ \citenamefont
  {Gopalakrishnan}}]{Karcher_Extended_2024a}%
  \BibitemOpen
  \bibfield  {author} {\bibinfo {author} {\bibfnamefont {J.~F.}\ \bibnamefont
  {Karcher}}, \bibinfo {author} {\bibfnamefont {R.}~\bibnamefont {Vasseur}},\
  and\ \bibinfo {author} {\bibfnamefont {S.}~\bibnamefont {Gopalakrishnan}},\
  }\bibfield  {title} {\bibinfo {title} {Extended critical phase in
  quasiperiodic quantum {{Hall}} systems},\ }\href
  {https://doi.org/10.1103/PhysRevB.109.064208} {\bibfield  {journal} {\bibinfo
   {journal} {Physical Review B}\ }\textbf {\bibinfo {volume} {109}},\ \bibinfo
  {pages} {064208} (\bibinfo {year} {2024})}\BibitemShut {NoStop}%
\bibitem [{\citenamefont {Wu}\ \emph {et~al.}(2022)\citenamefont {Wu},
  \citenamefont {Tang}, \citenamefont {Zhang},\ and\ \citenamefont
  {Zhang}}]{Wu_Quantized_2022}%
  \BibitemOpen
  \bibfield  {author} {\bibinfo {author} {\bibfnamefont {Y.-P.}\ \bibnamefont
  {Wu}}, \bibinfo {author} {\bibfnamefont {L.-Z.}\ \bibnamefont {Tang}},
  \bibinfo {author} {\bibfnamefont {G.-Q.}\ \bibnamefont {Zhang}},\ and\
  \bibinfo {author} {\bibfnamefont {D.-W.}\ \bibnamefont {Zhang}},\ }\bibfield
  {title} {\bibinfo {title} {Quantized topological {{Anderson-Thouless}}
  pump},\ }\href {https://doi.org/10.1103/PhysRevA.106.L051301} {\bibfield
  {journal} {\bibinfo  {journal} {Physical Review A}\ }\textbf {\bibinfo
  {volume} {106}},\ \bibinfo {pages} {L051301} (\bibinfo {year}
  {2022})}\BibitemShut {NoStop}%
\bibitem [{\citenamefont {Haldane}(1988)}]{Haldane_Model_1988}%
  \BibitemOpen
  \bibfield  {author} {\bibinfo {author} {\bibfnamefont {F.~D.~M.}\
  \bibnamefont {Haldane}},\ }\bibfield  {title} {\bibinfo {title} {Model for a
  {{Quantum Hall Effect}} without {{Landau Levels}}: {{Condensed-Matter
  Realization}} of the "{{Parity Anomaly}}"},\ }\href
  {https://doi.org/10.1103/PhysRevLett.61.2015} {\bibfield  {journal} {\bibinfo
   {journal} {Physical Review Letters}\ }\textbf {\bibinfo {volume} {61}},\
  \bibinfo {pages} {2015} (\bibinfo {year} {1988})}\BibitemShut {NoStop}%
\bibitem [{Note1()}]{Note1}%
  \BibitemOpen
  \bibinfo {note} {We have found that potentials with other values of $\beta $
  can similarly induce a reversal of Chern number, although generally following
  distinct mechanisms according to the microscopic details of the potential in
  each case. Likewise, a potential with an extra factor of $\cos (2\pi \beta
  (n+m))$, retaining the full 6-fold rotational symmetry of the honeycomb
  lattice, also drives a transition to $C=-1$. For simplicity of presentation,
  we focus on the potential of Eq. (\ref {Eq:Potential}) throughout this
  work.}\BibitemShut {Stop}%
\bibitem [{\citenamefont {Bianco}\ and\ \citenamefont
  {Resta}(2011)}]{Bianco_Mapping_2011}%
  \BibitemOpen
  \bibfield  {author} {\bibinfo {author} {\bibfnamefont {R.}~\bibnamefont
  {Bianco}}\ and\ \bibinfo {author} {\bibfnamefont {R.}~\bibnamefont {Resta}},\
  }\bibfield  {title} {\bibinfo {title} {Mapping topological order in
  coordinate space},\ }\href {https://doi.org/10.1103/PhysRevB.84.241106}
  {\bibfield  {journal} {\bibinfo  {journal} {Physical Review B}\ }\textbf
  {\bibinfo {volume} {84}},\ \bibinfo {pages} {241106} (\bibinfo {year}
  {2011})}\BibitemShut {NoStop}%
\bibitem [{SM()}]{SM}%
  \BibitemOpen
  \href@noop {} {}\bibinfo {note} {See Supplemental Material at [URL] for
  details on WPL state localisation, calculation of phase boundaries within the
  Born approximation, scaling analysis to verify the nature of different phases
  in the thermodynamic limit, and explicit calculation of the parameters $V_1$
  and $V_2$.}\BibitemShut {Stop}%
\bibitem [{\citenamefont {Zhao}\ \emph {et~al.}(2024)\citenamefont {Zhao},
  \citenamefont {Kang}, \citenamefont {Zhang}, \citenamefont {Kn{\"u}ppel},
  \citenamefont {Tao}, \citenamefont {Li}, \citenamefont {Tschirhart},
  \citenamefont {Redekop}, \citenamefont {Watanabe}, \citenamefont {Taniguchi},
  \citenamefont {Young}, \citenamefont {Shan},\ and\ \citenamefont
  {Mak}}]{Zhao_Realization_2024}%
  \BibitemOpen
  \bibfield  {author} {\bibinfo {author} {\bibfnamefont {W.}~\bibnamefont
  {Zhao}}, \bibinfo {author} {\bibfnamefont {K.}~\bibnamefont {Kang}}, \bibinfo
  {author} {\bibfnamefont {Y.}~\bibnamefont {Zhang}}, \bibinfo {author}
  {\bibfnamefont {P.}~\bibnamefont {Kn{\"u}ppel}}, \bibinfo {author}
  {\bibfnamefont {Z.}~\bibnamefont {Tao}}, \bibinfo {author} {\bibfnamefont
  {L.}~\bibnamefont {Li}}, \bibinfo {author} {\bibfnamefont {C.~L.}\
  \bibnamefont {Tschirhart}}, \bibinfo {author} {\bibfnamefont
  {E.}~\bibnamefont {Redekop}}, \bibinfo {author} {\bibfnamefont
  {K.}~\bibnamefont {Watanabe}}, \bibinfo {author} {\bibfnamefont
  {T.}~\bibnamefont {Taniguchi}}, \bibinfo {author} {\bibfnamefont {A.~F.}\
  \bibnamefont {Young}}, \bibinfo {author} {\bibfnamefont {J.}~\bibnamefont
  {Shan}},\ and\ \bibinfo {author} {\bibfnamefont {K.~F.}\ \bibnamefont
  {Mak}},\ }\bibfield  {title} {\bibinfo {title} {Realization of the {{Haldane
  Chern}} insulator in a moir\'e lattice},\ }\href
  {https://doi.org/10.1038/s41567-023-02284-0} {\bibfield  {journal} {\bibinfo
  {journal} {Nature Physics}\ }\textbf {\bibinfo {volume} {20}},\ \bibinfo
  {pages} {275} (\bibinfo {year} {2024})}\BibitemShut {NoStop}%
\bibitem [{\citenamefont {Wu}\ \emph {et~al.}(2019)\citenamefont {Wu},
  \citenamefont {Lovorn}, \citenamefont {Tutuc}, \citenamefont {Martin},\ and\
  \citenamefont {MacDonald}}]{Wu_Topological_2019}%
  \BibitemOpen
  \bibfield  {author} {\bibinfo {author} {\bibfnamefont {F.}~\bibnamefont
  {Wu}}, \bibinfo {author} {\bibfnamefont {T.}~\bibnamefont {Lovorn}}, \bibinfo
  {author} {\bibfnamefont {E.}~\bibnamefont {Tutuc}}, \bibinfo {author}
  {\bibfnamefont {I.}~\bibnamefont {Martin}},\ and\ \bibinfo {author}
  {\bibfnamefont {A.~H.}\ \bibnamefont {MacDonald}},\ }\bibfield  {title}
  {\bibinfo {title} {Topological {{Insulators}} in {{Twisted Transition Metal
  Dichalcogenide Homobilayers}}},\ }\href
  {https://doi.org/10.1103/PhysRevLett.122.086402} {\bibfield  {journal}
  {\bibinfo  {journal} {Physical Review Letters}\ }\textbf {\bibinfo {volume}
  {122}},\ \bibinfo {pages} {086402} (\bibinfo {year} {2019})}\BibitemShut
  {NoStop}%
\bibitem [{\citenamefont {Zhao}\ \emph {et~al.}(2021)\citenamefont {Zhao},
  \citenamefont {Xiao},\ and\ \citenamefont {Yao}}]{Zhao_Universal_2021}%
  \BibitemOpen
  \bibfield  {author} {\bibinfo {author} {\bibfnamefont {P.}~\bibnamefont
  {Zhao}}, \bibinfo {author} {\bibfnamefont {C.}~\bibnamefont {Xiao}},\ and\
  \bibinfo {author} {\bibfnamefont {W.}~\bibnamefont {Yao}},\ }\bibfield
  {title} {\bibinfo {title} {Universal superlattice potential for {{2D}}
  materials from twisted interface inside h-{{BN}} substrate},\ }\href
  {https://doi.org/10.1038/s41699-021-00221-4} {\bibfield  {journal} {\bibinfo
  {journal} {npj 2D Materials and Applications}\ }\textbf {\bibinfo {volume}
  {5}},\ \bibinfo {pages} {38} (\bibinfo {year} {2021})}\BibitemShut {NoStop}%
\bibitem [{\citenamefont {Wang}\ \emph {et~al.}(2025)\citenamefont {Wang},
  \citenamefont {Xu}, \citenamefont {Aronson}, \citenamefont {Bennett},
  \citenamefont {Paul}, \citenamefont {Crowley}, \citenamefont {Collignon},
  \citenamefont {Watanabe}, \citenamefont {Taniguchi}, \citenamefont {Ashoori},
  \citenamefont {Kaxiras}, \citenamefont {Zhang}, \citenamefont
  {{Jarillo-Herrero}},\ and\ \citenamefont {Yasuda}}]{Wang_Moire_2025}%
  \BibitemOpen
  \bibfield  {author} {\bibinfo {author} {\bibfnamefont {X.}~\bibnamefont
  {Wang}}, \bibinfo {author} {\bibfnamefont {C.}~\bibnamefont {Xu}}, \bibinfo
  {author} {\bibfnamefont {S.}~\bibnamefont {Aronson}}, \bibinfo {author}
  {\bibfnamefont {D.}~\bibnamefont {Bennett}}, \bibinfo {author} {\bibfnamefont
  {N.}~\bibnamefont {Paul}}, \bibinfo {author} {\bibfnamefont {P.~J.~D.}\
  \bibnamefont {Crowley}}, \bibinfo {author} {\bibfnamefont {C.}~\bibnamefont
  {Collignon}}, \bibinfo {author} {\bibfnamefont {K.}~\bibnamefont {Watanabe}},
  \bibinfo {author} {\bibfnamefont {T.}~\bibnamefont {Taniguchi}}, \bibinfo
  {author} {\bibfnamefont {R.}~\bibnamefont {Ashoori}}, \bibinfo {author}
  {\bibfnamefont {E.}~\bibnamefont {Kaxiras}}, \bibinfo {author} {\bibfnamefont
  {Y.}~\bibnamefont {Zhang}}, \bibinfo {author} {\bibfnamefont
  {P.}~\bibnamefont {{Jarillo-Herrero}}},\ and\ \bibinfo {author}
  {\bibfnamefont {K.}~\bibnamefont {Yasuda}},\ }\bibfield  {title} {\bibinfo
  {title} {Moir\'e band structure engineering using a twisted boron nitride
  substrate},\ }\href {https://doi.org/10.1038/s41467-024-55432-2} {\bibfield
  {journal} {\bibinfo  {journal} {Nature Communications}\ }\textbf {\bibinfo
  {volume} {16}},\ \bibinfo {pages} {178} (\bibinfo {year} {2025})}\BibitemShut
  {NoStop}%
\bibitem [{\citenamefont {Burgess}(2026)}]{Data}%
  \BibitemOpen
  \bibfield  {author} {\bibinfo {author} {\bibfnamefont {B.}~\bibnamefont
  {Burgess}},\ }\href {https://doi.org/10.17863/CAM.136855} {\bibinfo {title}
  {Research data and code supporting {``Reversing the Chern Number with
  Quasiperiodic Disorder"}}} (\bibinfo {year} {2026})\BibitemShut {NoStop}%
\bibitem [{\citenamefont {Aubry}\ and\ \citenamefont
  {Andr{\'e}}(1980)}]{Aubry_Analyticity_1980}%
  \BibitemOpen
  \bibfield  {author} {\bibinfo {author} {\bibfnamefont {S.}~\bibnamefont
  {Aubry}}\ and\ \bibinfo {author} {\bibfnamefont {G.}~\bibnamefont
  {Andr{\'e}}},\ }\bibfield  {title} {\bibinfo {title} {Analyticity breaking
  and {{Anderson}} localization in incommensurate lattices},\ }\href@noop {}
  {\bibfield  {journal} {\bibinfo  {journal} {Ann. Israel Phys. Soc}\ }\textbf
  {\bibinfo {volume} {3}},\ \bibinfo {pages} {18} (\bibinfo {year}
  {1980})}\BibitemShut {NoStop}%
\end{thebibliography}%

\appendix

\onecolumngrid
\vspace{\baselineskip}
\begin{center}
    \textbf{\large End Matter}
\end{center}
\vspace{0.5\baselineskip}
\twocolumngrid

\textit{Spectrum and localisation properties---}Here, we explicitly verify the spectral and localisation properties of the Haldane model under quasiperiodic and random disorder. We numerically diagonalise the Hamiltonian [Eq. (\ref{Eq:Hamiltonian})] for a finite system of $L \times L$ unit cells subject to open boundary conditions, with $L=30$ as in the main text. To quantify the spatial extent of the eigenstates, we calculate the inverse participation ratio (IPR) for each eigenstate $n$,
\begin{equation}
    \mathrm{IPR} = \sum_m |\psi_m^{(n)}|^4 \, ,
\end{equation}
where $\psi_m^{(n)}$ is the wavefunction amplitude at site $m$. For a state perfectly localized on a single site, $\mathrm{IPR}=1$, whereas for a state uniformly delocalised over the entire lattice, $\mathrm{IPR} \sim 1/L^2$, which vanishes in the thermodynamic limit. The localisation characteristics are conveniently captured by the fractal dimension $\gamma$:
\begin{equation}
    \gamma = -\frac{\ln(\mathrm{IPR})}{\ln L} \approx \begin{cases}
        2 \qquad & \text{extended states} \, , \\
        0 \qquad & \text{localised states} \, .
    \end{cases}
\end{equation}

\begin{figure*}[t]
    \centering
    \includegraphics[width=\textwidth]{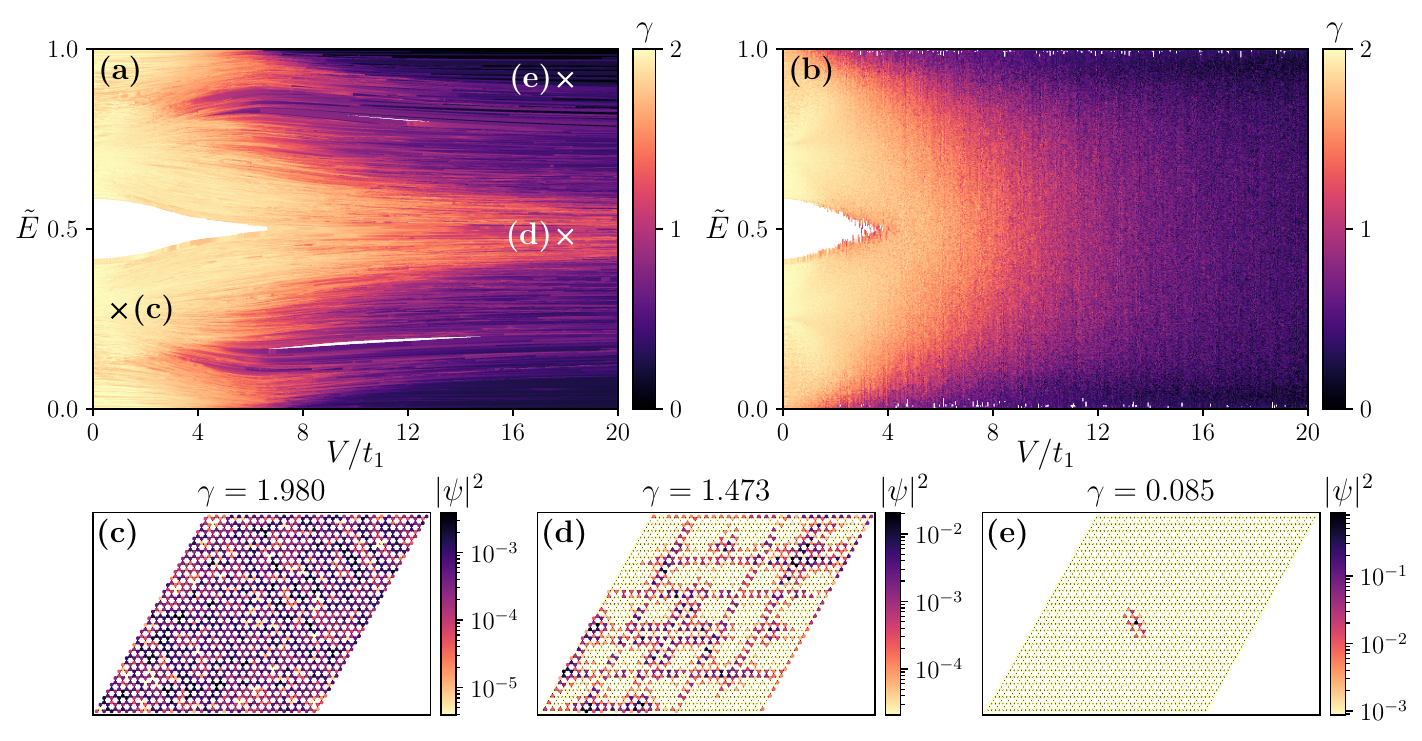}
    \caption{Spectrum of the Haldane model ($t_2/t_1 = 1.2$, $M=0$) for different strengths of (a) quasiperiodic and (b) random disorder. Energies are normalized as $\tilde{E}=(E-E_{\mathrm{min}})/(E_{\mathrm{max}}-E_{\mathrm{min}})$ to facilitate comparison across disorder amplitudes. The half-filling point, $\tilde{E}=0.5$, corresponds to $E\approx0$ (up to small variations depending on the specific disorder realisation). Points are coloured by fractal dimension $\gamma$. To avoid confusion between edge states, which are finite-size artefacts with $\gamma\sim 1$, and the bulk-supported WPL states, we exclude states with density $\sum_{\text{edge}} |\psi|^2 > 0.3$ within $2a$ of the boundary, where $a$ is the lattice constant. Crosses indicate representative eigenstates plotted in (c)--(e), illustrating typical extended, WPL, and localized states, respectively.}
    \label{Fig:Spectrum}
\end{figure*}

Figure~\ref{Fig:Spectrum} displays the eigenspectra of the two disorder types as a function of disorder amplitude $V$, alongside representative spatial probability densities $|\psi|^2$. In both cases, at weak disorder ($V \ll t_1$), all states are extended with $\gamma \approx 2$. As $V$ increases, all states in the randomly disordered system [Fig.~\ref{Fig:Spectrum}(b)] become localised with $\gamma\approx0$. In contrast, quasiperiodic disorder [Fig.~\ref{Fig:Spectrum}(a)] yields a mixed spectrum: localised states in the band tails coexist with extended states supported along WPLs around $E=0$, even for $V \gg t_1, t_2$. Because WPLs occupy only a fraction of the total 2D area, the finite-size calculations here yield intermediate values $0 < \gamma < 2$. In the Supplemental Material \cite{SM}, we quantitatively account for this deviation from the expected value $\gamma=2$ for extended states, confirming that the WPL states are genuinely extended across the 2D system rather than critically localized.

\newpage
\setcounter{secnumdepth}{2}

\renewcommand{\appendixname}{}

\renewcommand{\thesection}{S\arabic{section}}
\renewcommand{\thesubsection}{S\arabic{section}.\arabic{subsection}}
\renewcommand{\thefigure}{S\arabic{figure}}
\renewcommand{\theequation}{S\arabic{equation}}
\renewcommand{\thetable}{S\arabic{table}}

\setcounter{section}{0}
\setcounter{equation}{0}
\setcounter{figure}{0}
\setcounter{table}{0}

\onecolumngrid
\vspace{\baselineskip}
\begin{center}
    \textbf{\large Supplemental Material}
\end{center}
\vspace{0.5\baselineskip}
\twocolumngrid

\section{WPL State Localisation}\label{Sec:SM_WPL}

In this section, we explain the departure of the fractal dimension values $\gamma$ of the WPL states from the expected $\gamma=2$ for extended states in 2D. This results from the WPLs only covering a fraction of the sites of the system, and can be quantified as follows. A single line of constant $n$ can be described as an Aubry-André (AA) model \cite{Aubry_Analyticity_1980} in the $m$ direction, with Hamiltonian:
\begin{equation}
    H = \sum_m \left[ \left( t_2 \mathrm{e}^{\pm\mathrm{i}\phi} c_{m+1}^\dagger c_m + \mathrm{H.c.} \right) + V_n \cos(2\pi\beta m) c_m^\dagger c_m \right] \, .
\end{equation}
Here, $t_2$ is the relevant (intra-sublattice) hopping amplitude, and the sign of $\phi$ (determined by the sublattice) does not affect localisation. The potential amplitude is $V_n=V\cos(2\pi\beta n)$ (we set the phase offsets $\theta_{1,2}=0$ without loss of generality). This line is a WPL with extended states if $|V_n| < V_c$, where $V_c=2t_2$ is the critical potential for localisation determined by the self-dual point of the AA model \cite{Aubry_Analyticity_1980}. Since $\beta$ is irrational, the phases $2\pi\beta n \: (\mathrm{mod} \,2\pi)$ are dense and uniform on $[0, 2\pi)$. The probability $p$ that line $n$ is a WPL is therefore simply the fraction of the cosine domain satisfying $|V_n| < V_c$, which yields:
\begin{equation}
    p = \frac{2}{\pi}\arcsin\left(\frac{2t_2}{V}\right) \, .
\end{equation}
The probability that any site $(n,m)$ lies on a WPL is then $2p-p^2$, with the $p^2$ term correcting for double-counting of the intersections of WPLs. A state delocalised over all WPL sites will have $\mathrm{IPR}\sim1/(2p-p^2)L^2$, and therefore fractal dimension $\gamma=2 + \log(2p-p^2)/\log(L)$. Substituting parameters $t_2/t_1=1.2$, $V/t_1=18$ and $L=30$, we find $\gamma\approx 1.44$, consistent with the range of values $1.3 \lesssim \gamma \lesssim 1.6$ seen for WPL states in Fig. 5(a) of the main text. This demonstrates that the WPL states are genuinely extended in 2D with $\gamma=2$, up to this finite-size correction $\log(2p-p^2)/\log(L)$ which vanishes in the limit $L \to \infty$.
\section{Born Approximation Phase Boundaries}

In this section, we give details of the calculations to obtain the TAI-BI phase boundaries, following the general procedure of Ref. \cite{Groth_Theory_2009}. The effect of disorder is captured by introducing the self-energy $\Sigma$, defined by
\begin{equation}\label{Eq:Self_Energy}
    \overline{G^{-1}} = G_0^{-1}-\Sigma\, ,
\end{equation}
where $G_0=(E-H_0+\mathrm{i}0^+)^{-1}$ is the Green's function of the clean Haldane model with Hamiltonian $H_0$, $G=(E-H_0-V+\mathrm{i}0^+)^{-1}$ is the Green's function for the system with a specific disorder configuration $V$, and $\overline{\cdots}$ indicates averaging over disorder configurations. Within the Born approximation, $\Sigma$ is given by
\begin{equation}\label{Eq:SigmaBorn}
    \Sigma=\overline{V G_0 V} \, .
\end{equation}
We work in the momentum representation, where $\Sigma$, $V$ and $G_0$ are $2\times 2$ matrices in the sublattice basis. For the quasiperiodic potential, Eq. (\ref{Eq:SigmaBorn}) becomes
\begin{equation}
    \Sigma(\mathbf{k}) = \sum_{\mathbf{q}}|V_{\mathbf{q}}|^2 
    \begin{pmatrix}
        1 & 0 \\ 
        0 & \mathrm{e}^{\mathrm{i}\Delta \phi_\mathbf{q}}
    \end{pmatrix}
    G_0(\mathbf{k}+\mathbf{q}) 
    \begin{pmatrix}
        1 & 0 \\ 
        0 & \mathrm{e}^{-\mathrm{i}\Delta \phi_\mathbf{q}}
    \end{pmatrix} \, ,
\end{equation}
where $V_\mathbf{q}$ are the amplitudes of Fourier components $\mathbf{q}$ of the potential, and $\Delta \phi_\mathbf{q}=\mathbf{q}\cdot (\mathbf{r}_B - \mathbf{r}_A)$ is the phase shift between $A$ and $B$ sites for that component. Rewriting the potential as:
\begin{equation}
    \begin{aligned}
        V_i & =V\cos(2\pi\beta n)\cos(2\pi\beta m) \\
        & = V \cos(\beta \mathbf{G}_1 \cdot\mathbf{r}_i) \cos(\beta \mathbf{G}_2 \cdot\mathbf{r}_i) \\ 
        & = \frac{V}{2} \left[ \cos(\beta (\mathbf{G}_1+\mathbf{G}_2) \cdot\mathbf{r}_i)
        + \cos(\beta (\mathbf{G}_1-\mathbf{G}_2) \cdot\mathbf{r}_i) \right] \, ,
    \end{aligned}
\end{equation}
where $\mathbf{G}_{1,2}$ are the reciprocal vectors of the honeycomb lattice, we can read off that $\mathbf{q}\in \{ \pm \beta (\mathbf{G}_1 \pm \mathbf{G}_2 ) \}$, and $|V_\mathbf{q}|^2=V^2/16$ for all components. Including the phase offsets $\theta_{1,2}$ does not affect $|V_\mathbf{q}|$. 

The phase transition between the CI and BI phases occurs when the band gap of the system closes, that is $H_0+\Sigma=0$ (for all matrix components) at some point in $\mathbf{k}$-space. We expand $\Sigma$ in terms of Pauli matrices as
\begin{equation}
    \Sigma = \Sigma_0 \mathbbm{1}_2 
    + \sum_{i\in \{x, y, z\}} \Sigma_i \sigma_i \, .
\end{equation}
In practice, we find that the off-diagonal parts $\Sigma_{x,y}$ are negligibly small, and so the gap closes at the same position as in the clean case, at the $K$ point of the BZ. The $\Sigma_0$ term also vanishes identically. At potential magnitude $V$, the critical value of $M$ for the phase transition is therefore given by the condition
\begin{equation}\label{Eq:General_Phase_Boundary}
    M - M_0 + \Sigma_z(\mathbf{k}=\mathbf{k}_K,M,V)=0 \, ,
\end{equation}
where $M_0=3\sqrt{3}t_2$ is the critical value of $M$ in the clean system. Solving Eq. (\ref{Eq:General_Phase_Boundary}) numerically for $M$, we obtain the phase boundary plotted in Fig. 2(a) of the main text.

For random disorder, $V$ is independent of $\mathbf{k}$, and therefore so is $\Sigma$. This enables us to use the self-consistent Born approximation (SCBA), which amounts to solving Eq. (\ref{Eq:SigmaBorn}) with the bare $G_0$ replaced by the Green's function including the self-energy:
\begin{equation}\label{Eq:SCBA}
    \Sigma = \frac{V^2}{3} \int_{\mathrm{BZ}} \frac{\mathrm{d}^2 \mathbf{k}}{A_{\mathrm{BZ}}} \sum_{\alpha\in \{A, B\}} P_\alpha (G_0^{-1}-\Sigma)^{-1} P_\alpha \, ,
\end{equation}
where $V^2/3$ is the variance of the disorder, the integral is over the BZ with area $A_{\mathrm{BZ}}$, and the projectors $P_\alpha$ onto sublattice $\alpha$ are included due to the absence of correlations between $A$ and $B$ sites. The phase boundary in Fig. 2(b) is obtained by inserting the result of Eq. (\ref{Eq:SCBA}) into Eq. (\ref{Eq:General_Phase_Boundary}) and solving numerically for $M$.

The Born approximation becomes less accurate as $V$ increases due to the neglection of higher-order terms in the expression for $\Sigma$. This approach also fails to capture phase transitions due to localisation; the self energy here is calculated from the disorder average of a single Green's function [Eq. (\ref{Eq:Self_Energy})], whereas properties which relate to localisation, such as the conductivity, involve the disorder average of a \textit{pair} of Green's functions.
\section{Nature of Different Phases}\label{Sec:SM_Scaling}

In this section, we verify that the NCI is a genuine Chern insulator phase, characterised by nonzero bulk energy gap $\Delta E$ and integer $\overline{c}$ in the limit of $L\to\infty$. To do so, we investigate the variation of $\Delta E$ and $\overline{c}$ with increasing $L=20, 24,\ldots,80$. We compare $V/t_1=4,\,8,\,16$, which are typical values of the CI, NCI and AI+WPL phases respectively.

$\Delta E$ is calculated as the energy gap at half-filling after projecting out edge states. For $V/t_1=4$, we define edge states as those with total density $\sum|\psi|^2>0.3$ within $2a$ of the edge, where $a$ is the lattice constant. For $V/t_1=8,\,16$ where WPLs become significant, we instead define edge states as those with total density $\sum|\psi|^2>0.5$ within $5a$ of the edge; this is because WPLs are spaced by approximately $5a$, so the edge \textit{of the network of WPLs} may be anywhere within $5a$ of the edge \textit{of the system}. For all $L$, $\overline{c}$ is calculated as the average of $c(\mathbf{r})$ over the central $6 \times 6$ unit cells of the system, the same as in the main text.

\begin{figure}[ht]
    \centering
    \includegraphics[width=0.95\linewidth]{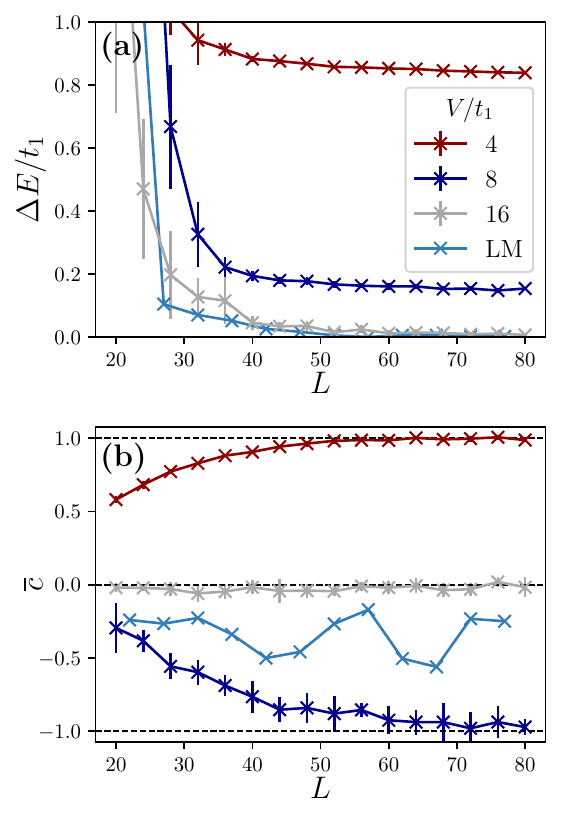}
    \caption{Variation of (a) bulk energy gap $\Delta E$ and (b) bulk-averaged Chern marker $\overline{c}$ with system size $L$. System parameters were $t_2/t_1=1.5$ (as in Fig. 2 of the main text), $M=0$, and $V$ as detailed in the legend in (a) (applying to both plots). Each point was obtained as the mean of 10 disorder configurations, with the standard deviation of results indicated by the error bars. For the ladder model (labelled LM), the parameters were $(V_1/t_1, V_2/t_1)=(6, 5)$; since this model is periodic, no averaging over configurations was required.}
    \label{Fig:Scaling}
\end{figure}

The results are plotted in Fig. \ref{Fig:Scaling}. For both the CI and NCI, we see a clear plateau in $\Delta E$ as $L$ increases, indicating a band gap in the thermodynamic limit; likewise, the $\overline{c}$ values tend towards $+1$ and $-1$ respectively. Conversely, for the trivial AI+WPL phase, $\Delta E$ tends to zero as $L$ increases, and $\overline{c}$ remains near-zero throughout. 

We perform the same analysis for the ladder network model, labelled LM in Fig. \ref{Fig:Scaling}. Bulk states are defined in the same way as the $V/t_1=8,\,16$ cases. As a representative example, we choose parameters $(V_1/t_1, V_2/t_1)=(6, 5)$, within the negative $\overline{c}$ region of Fig. 4. Here we find a vanishing $\Delta E$, and $\overline{c}$ failing to converge to an integer, indicating a metallic phase. This persists across much of the parameter space of Fig. 4; the exception is for large $(V_1, V_2)$ where a gap opens, signalling a transition to a trivial band insulator phase. As described in the main text, although $\overline{c}$ is not quantised in the metallic phase, we can still use the sign of $\overline{c}$ to qualitatively understand the topology in the full system. However, this variance between the metallic ladder phase and the insulating NCI underlines the simplifications built into the ladder model, cautioning that one should not expect precise quantitative agreement of, for example, phase boundaries.

\section{Calculation of Effective Potentials}

\begin{figure}[ht]
    \centering
    \includegraphics[width=\linewidth]{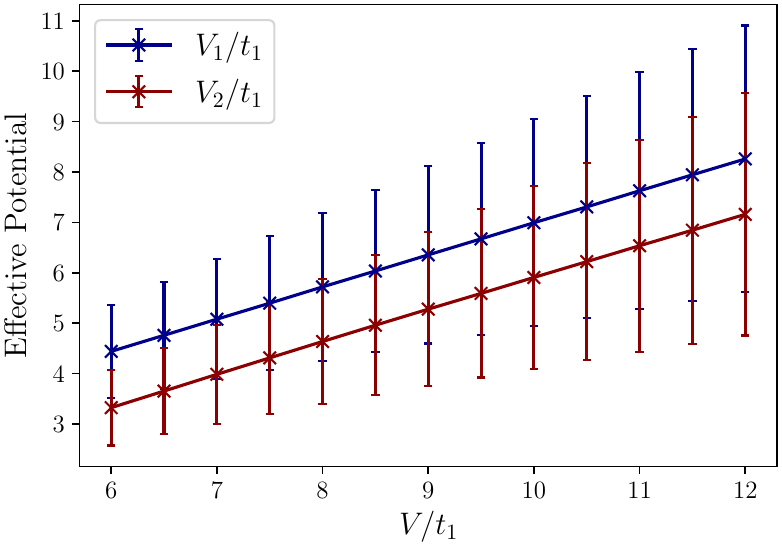}
    \caption{Calculated mean effective potentials $V_1$ and $V_2$, with standard deviations indicated by the error bars.}
    \label{Fig:V1V2}
\end{figure}

In this section, we detail the calculation of the effective parameters $V_1$ and $V_2$. As in Sec. \ref{Sec:SM_WPL}, WPLs are classified as lines satisfying either $V \cos(2\pi\beta n + \theta_1)<V_c$ or $V \cos(2\pi\beta m + \theta_2)<V_c$, where $V_c=2t_2$ is the critical localisation potential of the Aubry-André model. The values of $V_1$ and $V_2$ are given by the mean potential magnitudes of the relevant sites adjacent to the WPLs (see Fig. 3 in the main text). We use a $L\times L$ system with $L=1000$, with randomly chosen potential phase offsets $\theta_{1,2}$ (for such large $L$, the statistical variation for different $\theta_{1,2}$ is negligibly small). The calculated values are plotted in Fig. \ref{Fig:V1V2} for potential strengths $V$ relevant to the NCI phase. Error bars show the standard deviation of values across the finite system. Since the set of sites which are classified as WPLs changes as $V$ increases, $V_1$ and $V_2$ are not quite (albeit very nearly) linear with $V$. 

From the quasiperiodic phase diagram [main text, Fig. 2(a)], the NCI phase is present for potential strengths $6 \lesssim V/t_1 \lesssim10$, for which the calculations here [Fig. \ref{Fig:V1V2}] give a range $(4.4, 3.3) \lesssim (V_1/t_1, V_2/t_1) \lesssim(7.0, 5.9)$. This range is reasonably consistent with the values $(V_1, V_2)$ yielding a negative Chern marker in the ladder network phase diagram [main text, Fig. 4]. This consistency validates our use of this effective model to understand the topological character of the NCI phase in the full quasiperiodic system, implying that the negative $\overline{c}$ shares the same physical origin in both cases. As discussed in Sec. \ref{Sec:SM_Scaling}, we do not expect exact quantitative agreement due to the approximations made in the construction of the ladder model, as well as the uncertainty in inferring the exact phase boundaries from Fig. 2.

\end{document}